\documentclass[reprint,3p,sort&compress,times,onecolumn]{elsarticle}
\usepackage{amssymb}
\usepackage{amsmath}
\usepackage{bm}
\usepackage{gensymb}
\usepackage{stfloats}
\usepackage{bm}
\usepackage{pifont}
\usepackage{graphicx}
\usepackage{subfigure}
\usepackage{float}
\usepackage{booktabs}
\usepackage{caption}
\usepackage[colorlinks=true,linkcolor=blue,citecolor=blue,anchorcolor=blue,urlcolor=blue]{hyperref}

    \journal{******}
\usepackage{lineno}

\begin{document}

\begin{frontmatter}

\title{Thermodynamically consistent non-isothermal phase-field modelling of elastocaloric effect: indirect \textit{vs} direct method}

\author[rvt]{Wei Tang}

\author[rvt,rvg]{Qihua Gong}

\author[rvt]{Min Yi\corref{cor1}}
\ead{yimin@nuaa.edu.cn}
\cortext[cor1]{Corresponding author}

\author[els]{Bai-Xiang Xu}

\author[dres]{Long-Qing Chen}



\address[rvt]{State Key Laboratory of Mechanics and Control 
for Aerospace Structures \& Key Laboratory for Intelligent Nano Materials and Devices of Ministry of Education \& Institute for Frontier Science \& College of Aerospace Engineering, Nanjing University of Aeronautics and Astronautics (NUAA), Nanjing 210016, China}

\address[rvg]{MIIT Key Lab of Aerospace Information Materials and Physics \& College of Physics, Nanjing University of Aeronautics and Astronautics (NUAA), Nanjing
211106, China}

\address[els]{Mechanics of Functional Materials Division, Institute of Materials Science, Technische Universit$\ddot{a}$t Darmstadt, Darmstadt 64287, Germany}

\address[dres]{Department of Materials Science and Engineering and Materials Research Institute, The Pennsylvania State University, PA 16803, USA}

\begin{abstract}
Modelling elastocaloric effect (eCE) is crucial for the design of environmentally friendly and energy-efficient eCE based solid-state cooling devices. Here, a thermodynamically consistent non-isothermal phase-field model (PFM) coupling martensitic transformation with mechanics and heat transfer is developed and applied for simulating eCE. The model is derived from a thermodynamic framework which invokes the microforce theory and Coleman--Noll procedure. To avoid the numerical issue related to the non-differentiable energy barrier function across the transition point, the austenite-martensite transition energy barrier in PFM is constructed as a smooth function of temperature. Both the indirect method using isothermal PFM with Maxwell relations and the direct method using non-isothermal PFM are applied to calculate the elastocaloric properties. The former is capable of calculating both isothermal entropy change and adiabatic temperature change ($\Delta T_{\text{ad}}$), but induces high computation cost. The latter is computationally efficient, but only yields $\Delta T_{\text{ad}}$. In a model Mn-22Cu alloy, the maximum $\Delta T_{\text{ad}}$ ($\Delta T_{\text{ad}}^{\text{max}}$) under a compressive stress of 100 MPa is calculated as 9.5 and 8.5~K in single crystal (3.5 and 3.8~K in polycrystal) from the indirect and direct method, respectively. It is found that the discrepancy of $\Delta T_{\text{ad}}^{\text{max}}$ by indirect and direct method is within 10\% at stress less than 150 MPa, confirming the feasibility of both methods in evaluating eCE at low stress. However, at higher stress, $\Delta T_{\text{ad}}^{\text{max}}$ obtained from the indirect method is notably larger than that from the direct one. This is mainly attributed to that in the non-isothermal PFM simulations, the relatively large temperature increase at high stress could in turn hamper the austenite-martensite transition and thus finally yield a lower $\Delta T_{\text{ad}}$. The results demonstrate the developed PFM herein, combined with both indirect and direct method for eCE calculations, as a practicable toolkit for the computational design of elastocaloric devices.
\end{abstract}

\begin{keyword}
Elastocaloric effect, Phase-field model, Shape memory alloys, Martensitic transformation, Adiabatic temperature change

\end{keyword}

\end{frontmatter}

\section{Introduction}
The solid-state cooling technology features itself as an environmentally friendly and efficient alternative to the traditional vapor compression refrigeration, thus helping move towards carbon neutrality.
Typical solid-state cooling by  magnetocaloric~\cite{wada2001giant,phan2007review,gschneidnerjr2005recent}, electrocaloric~\cite{mischenko2006giant,scott2011electrocaloric},
barocaloric
~\cite{manosa2010giant,lloveras2015giant} and elastocaloric~\cite{bonnot2008elastocaloric,tuvsek2015elastocaloricni,manosa2013large,tuvsek2015elastocaloric} effects has been widely investigated, among which elastocaloric cooling stands out with the large temperature change~\cite{tuvsek2015elastocaloricni,cui2012demonstration}, large working temperature window~\cite{manosa2013large}, and excellent coefficient of performance~\cite{ovzbolt2014electrocaloric,qian2016review}. The elastocaloric cooling is realized by means of elastocaloric effect (eCE), which originates from the latent heat associated with martensitic transformation (MT) in shape memory alloys (SMAs).
Upon loading, the exothermic austenite-martensite transformation (also called the conventional MT) would cause a temperature increase in the adiabatic process. Upon unloading, the endothermic martensite-austenite transformation (also called the inverse MT) occurs, and a rapid drop in temperature arises~\cite{bonnot2008elastocaloric}. 
The eCE can be quantified by the adiabatic temperature change ($\Delta T_{\text{ad}}$) or the isothermal entropy change ($\Delta S_{\text{iso}}$). 

Modelling eCE plays an important role in the computational design of elastocaloric devices and can be an essential complement to experiments.
In general, phenomenological constitutive models and phase-field model~\cite{chen2002phase,chen2022classical} (PFM) are utilized to calculate the eCE in SMAs.
The phenomenological Tanaka-type model~\cite{tanaka1986thermomechanical,ossmer2014evolution,krevet2013evolution,tuvsek2016understanding,tuvsek2015elastocaloric,luo2017modeling} has advantages in the description of temperature change and stress-strain behavior. 
Very good agreement between the measured and calculated values of $\Delta T_{\text{ad}}$ is obtained. Analogously, the phase transformation kinetic model~\cite{qian2015thermodynamics,qian2017mechanism,qian2017numerical}
and crystal plasticity-based constitutive model~\cite{yu2020modeling,zhou2020modeling} are also proposed to predict the eCE. These phenomenological constitutive models can be used to obtain the macroscopic elastocaloric properties, but have difficulties in simulating the spatial and temporal evolution of microstructure details during the MT and thus are hardly applicable to the optimization of eCE by microstructure engineering.

eCE in SMAs is intrinsically ascribed to MT, which is a first-order phase transformation and can be induced by external stress/strain field or by temperature field. The PFM developed from Landau's theory of phase transformation is widely used to simulate the stress- and temperature-induced MT~\cite{chen2004phase}. Wang et al.~\cite{wang1997three} built a three-dimensional PFM of MT, taking into account the transformation-induced strain, which comprehensively described a generic cubic-to-tetragonal MT. In the PFM, each of the martensitic variants is described by an order parameter $\eta_I$ ($I = 1, 2, ... , n$ with $n$ as the total number of various crystallographically equivalent martensitic variants), and the evolution of each of variants is governed by the Ginzburg-Landau equations. Several studies also considered hexagonal-to-orthorhombic~\cite{wen2000phase}, cubic-to-tetragonal~\cite{seol2003cubic}, and tetragonal-to-monoclinic~\cite{mamivand2013phase} transitions.
Over the last decades, the PFM for MT has been constantly improved to include/study more physical phenomena in SMAs. For example, the martensitic reorientation, the temperature-induced transformation, and the stress-induced MT have been of high interests. 
Levitas et al.~\cite{levitas2002three,levitas2002three2,levitas2003three} constructed a thermodynamically consistent PFM for transformations between austenite and martensitic variants and martensitic reorientation. 
Since then the phase-field (PF) theory is extended for the cases of surface stresses~\cite{levitas2010surface,levitas2013thermodynamically}, large strain~\cite{levitas2009displacive,levitas2013phase}, and eCE~\cite{hou2019fatigue,cisse2020elastocaloric,cisse2020asymmetric,CISSE2021109898}. The research that explores eCE by PFM has recently emerged, owing to the rise of efficient solid-state cooling technology by eCE as well as the PFM's advantage in simulating the evolution of microstructure and temperature  during the MT process.

In 2015, Levitas et al.~\cite{levitas2015multiphase} proposed a multiphase phase-field theory for temperature- and stress-induced MT, which allows for a presence of the third phase at the interface between the two other phases. Then, Cui et al.~\cite{cui2017three} employed a non-isothermal PFM to study the stress- and temperature-induced MT, as well as the corresponding latent heat in SMAs. However, the discontinuous piecewise energy barrier between austenite and martensite is adverse to the calculation of $\Delta S_{\text{iso}}$ or $\Delta T_{\text{ad}}$, which possibly makes the Maxwell relations based eCE analysis problematic around the transition point.
In real materials, the austenite-martensite transition energy barrier varies continuously and increases with temperature in the whole temperature range. Furthermore, Sun et al.~\cite{sun2018phase,sun2019non} modified this energy barrier as a continuous piecewise function of temperature in PFM and discussed the effect of grain size, crystal orientation, and loading rate on $\Delta T_{\text{ad}}$. 
Ciss{\'e} et al.~\cite{cisse2020elastocaloric} introduced an exponential function to express the energy barrier and computed qualitatively the eCE in CuAlBe, including $\Delta T_{\text{ad}}$ calculated by the direct method, coefficient of performance and cyclic deformation. Their energy barrier equations are continuous but not differentiable at the transition point (see Fig.~\ref{f-energy} for more details).
In addition, Xu et al.~\cite{xu2020phase,xu2021phase} constructed a similar PFM by introducing an extra grain boundary energy to investigate the grain size dependent super-elasticity and eCE using the direct method in nanocrystalline NiTi SMAs. A mesoscale model~\cite{wendler2017mesoscale} based on the Müller-Achenbach-Seelecke theory without PF order parameters was also developed to describe the  evolution of local temperature and strain in SMA films during elastocaloric cycling. Meanwhile, the latent heat effect~\cite{cui2017three,sun2019non}, size effect~\cite{sun2018phase,hou2021effect,xu2020phase},  plasticity~\cite{cisse2020elastocaloric,cisse2020asymmetric} , and microstructure design~\cite{xu2021phase,CISSE2021109898} are considered to regulate eCE. 
Nevertheless, all the above mentioned works are focused on the application of PF simulations to directly evaluate $\Delta T_{\text{ad}}$ during loading and unloading by solving the heat-transfer equation. 

In general, the indirect and direct methods are two basic approaches to calculate the key parameter $\Delta T_{\text{ad}}$~\cite{moya2020caloric}.
Directly measuring the (adiabatic) temperature change distribution by thermography without thermodynamic calculations is called the direct method.
The difficulty of ensuring adiabatic conditions usually leads to the inevitable heat loss and thus the minor $\Delta T_{\text{ad}}$ calculated by the direct method.
Hence, researchers also adopt the indirect method~\cite{yuan2019elastocaloric,chen2019giant,hou2018ultra} to calculate $\Delta S_{\text{iso}}$ and $\Delta T_{\text{ad}}$ by means of the Maxwell relations (Eq.~\ref{eq1}) or Clausius-Clapeyron equation. $\Delta S_{\text{iso}}$ and $\Delta T_{\text{ad}}$ can be calculated from the average stress-strain responses at each temperature or average strain-temperature responses at each stress. 
The Maxwell relations for eCE evaluation read as
\begin{equation}
  \rho  \left( \frac{\partial S}{\partial\sigma}\right)_T = \left( \frac{\partial \varepsilon}{\partial T}\right)_\sigma ~ \text{or} \quad \rho  \left( \frac{\partial S}{\partial\varepsilon}\right)_T = -\left( \frac{\partial \sigma}{\partial T}\right)_\varepsilon .\label{eq1}
\end{equation}
Using Eq.~\ref{eq1}, $\Delta S_{\text{iso}}$ and $\Delta T_{\text{ad}}$ between the initial and final stress are expressed as~\cite{bonnot2008elastocaloric}
\begin{equation}
    \Delta S_{\text{iso}} (0 \to \sigma)= \int_0^{\sigma} \frac{1}{\rho} \left(\frac{\partial \varepsilon}{\partial T}\right)_\sigma \text{d}\sigma , \label{eq26}
\end{equation}
\begin{equation}
    \Delta T_{\text{ad}} (0 \to \sigma)=- \int_0^\sigma \frac{T}{ \rho c} \left(\frac{\partial \varepsilon}{\partial T}\right)_\sigma \text{d}\sigma \label{eq27}
\end{equation}
where $S$ is entropy, $T$ temperature, $\varepsilon$  strain, $\sigma$ stress, $\rho$ material's density, and $c$ specific heat.

The indirect method is initially extensively utilized to calculate the magnetocaloric effect~\cite{phan2007review,gschneidnerjr2005recent}. 
Bonnot et al.~\cite{bonnot2008elastocaloric} first introduced the Maxwell relations (Eq.~\ref{eq1}) to evaluate $\Delta S_{\text{iso}}$ associated with the MT in Cu-Zn-Al single crystal and found that the indirectly calculated $\Delta S_{\text{iso}}$ agrees well with the directly measured value. This work has brought about the widespread use of the indirect method for evaluating eCE experimentally. 
Chen et al.~\cite{chen2019giant} found the directly measured $\Delta T_{\text{ad}}$ in nanocrystalline Ti-Ni-Cu SMA is consistent with $\Delta T_{\text{ad}}$ from the indirect method. 
However, Pataky et al.~\cite{pataky2015elastocaloric} demonstrated that the directly measured $\Delta T_{\text{ad}}$ is much lower than the temperature change calculated by the Maxwell or Clausius-Clapeyron relations. Similarly, Qian et al.~\cite{qian2016elastocaloric} measured the eCE in CuAlZn and CuAlMn SMAs under compression, and showed that $\Delta T_{\text{ad}}$ estimated by the indirect method is nearly three times as large as the directly measured one. Experimental results indicate the discrepancy in $\Delta T_{\text{ad}}$ from the indirect and direct method, resulting in some misunderstandings and disputes about the eCE in the same material, similar observations have been reported for electrocaloric effect of ferroelectric materials  ~\cite{grunebohm2018origins}.
It implies that if the stress-strain curves across the transition point is not measured with a sufficiently small incremental step of temperature, the indirect method may lead to unrealistic $\Delta S_{\text{iso}}$ and $\Delta T_{\text{ad}}$ owing to the possibly incorrect estimation of $\partial\varepsilon/\partial T$~\cite{tuvsek2016understanding,moya2020caloric}.
In short, the literature review indicates the calculation method and its effectiveness as a key issue for evaluating eCE by PF simulations. The indirect method has the merit of calculating both $\Delta S_{\text{iso}}$ and $\Delta T_{\text{ad}}$, whereas the direct method only attains $\Delta T_{\text{ad}}$.
However, the indirect method based on the Maxwell relations, which is widely adopted in experimental studies of eCE, is relatively unexplored in terms of PF simulations. Whether the eCE evaluated by the indirect and direct method using PFM is consistent with each other remains yet unknown. 

In this work, we aim to develop a thermodynamically consistent non-isothermal PFM coupling MT with mechanics and heat transfer to evaluate eCE using indirect and direct method. The model is derived from a thermodynamic framework which invokes the microforce theory and the Coleman--Noll procedure. In order to account for the possibly non-sharp transition in real materials and avoid the problematic calculation due to the non-differentiable energy barrier function across the transformation temperature, the austenite-martensite transition energy barrier in PFM is introduced as a smooth function of temperature. The PFM details and their derivation accomplished by a systematic use of thermodynamic principles are presented in Sect.~\ref{sec2} for both improper and proper MT.
Sect.~\ref{sec-application} gives an application of non-isothermal PFM to Mn-22Cu alloy.
In Sect.~\ref{sec3}, the PFM for both single crystal and polycrystal is numerically implemented by the finite element method. 
In Sect.~\ref{sec4}, benchmark simulations for reproducing the ferroelastic behavior and elastocaloric cycle are performed to verify the PFM.  
In addition, the elastocaloric properties in Mn-22Cu SMAs are obtained through the indirect and direct method, and the comparison of $\Delta T_{\text{ad}}$ calculated from these two methods are discussed. 
Sect.~\ref{sec5} gives the conclusive summary.

\section{Thermodynamically consistent phase-field model}\label{sec2}
The phase-field method emerges as a powerful tool for modelling  microstructure evolution and predicting eCE in elastocaloric materials. It describes a microstructure using a set of conserved or non-conserved order parameters (OPs) that are continuous across the smooth interfacial regions~\cite{anderson1998diffuse}, which could implicitly track the positions of interfaces. For deriving the partial differential equations for the non-isothermal PFM, we consider a closed system of volume $\Omega$ in which a pure material undergoes a first-order phase transformation between austenite and martensite. Noticeably, the MT could be divided into improper and proper MT.
The improper MT is characterized by the displacements of atoms within a unit cell and is described by soft optical displacement modes (e.g., ceramic materials), where strain is generated as a secondary effect. The proper MT is a homogeneous stress-free (eigen) strain that characterizes a change in crystal lattice parameters~\cite{Khachaturyan1997Three,artemev2001three}, e.g., fcc to bcc MT in Fe alloys. Therefore, the proper and improper MT in SMAs should be distinguished and modelled by PFM using different OPs. 
In 1995, Kartha et al~\cite{Kartha1995Disorder} proposed a phase field theory for the proper MT by using the strain tensor, which is based on a straightforward approach that uses the Helmholtz free energy defined by the OPs of eigen strain $\varepsilon_{ij}^0$. In addition, the non-conserved OPs $\eta_I$ are also chosen to distinguish various phases in SMAs regardless of the improper or proper MT~\cite{cui2017three}.

In this work, non-conserved OPs $\chi$ is chosen to describe various phases during MT in SMAs. For the improper MT, $\chi$ is $\eta_I$ which is chosen to distinguish various phases: $\eta_I=0$ represents austenite and $\eta_I=1$ represents the $I^{\text{th}}$ martensitic variant. For the proper MT, the OPs $\chi$ is the eigen strain $\varepsilon_{ij}^0$, which can directly represent the martensitic variants and austenite.
The following subsections present the complete thermodynamic consistent derivation that involves balance law, Coleman--Noll analysis, constitutive relations and evolution equations, and Helmholtz free energy.

\subsection{Balance law}

\begin{itemize}
\item Balance of linear momentum

For the body $\Omega$ with a boundary $\partial\Omega$, the quasi-static mechanical equilibrium equation is described by
\begin{equation}
\sigma_{ji,j} + b_i= 0 \quad \text{in}~\Omega ~, \label{eq-mecheq}
\end{equation}
\begin{equation}
u_i=\hat{u}_i \quad \text{on}~\partial\Omega_u, \quad \sigma_{ji}n_{j}=\hat{t}_i \quad \text{on}~\partial\Omega_{\sigma}
\end{equation}
where $\sigma_{ji}$ is the Cauchy stress tensor and $b_i$ is the body force. Here, we assume $b_i=0$. The Latin indices ($i$, $j$) run over the range of 1--3. $\hat{u}_i$ is the displacement prescribed on the boundary $\partial\Omega_u$, $n_j$ is the outward surface unit vector, and $\hat{t}_i$ is the surface traction on the boundary $\partial\Omega_{\sigma}$~\cite{yi2016real}.

\item Balance of angular momentum

The stress tensor is symmetric, i.e.,
\begin{equation}
\sigma_{ij}=\sigma_{ji} . \label{eq-angular-m}
\end{equation}

\item Balance of microforce associated with OPs $\chi$

Based on the Gurtin's microforce theory~\cite{gurtin1996generalized}, we assume that there exist a set of forces that accounts for the phase transition dynamics. These forces are called microforces because they are involved with the local transformation of the material, rather than the macroscopic movements. In this work, the microforce is associated with the non-conserved OPs $\chi$, describing the phase transformation between austenite and martensite.
The relevant microforce definitions are as follows.

$\xi^{\chi}$: the microstress, \\
$\xi^{\chi} n_i$: the surface microforce with $n_i$ as the unit outer normal to $\partial\Omega$, \\
$\zeta^{\chi}$: the internal microforce, \\
$\zeta^{\chi \text{ex}}$: the external microforce.

The balance of microforce associated with the phase transformation is written as
\begin{equation}
\int_{\partial\Omega} \xi^{\chi} n_i \text{d}A + \int_{\Omega}\zeta^{\chi} \text{d}V + \int_\Omega \zeta^{\chi \text{ex}} \text{d}V=0.
\end{equation}

By means of the Gauss law, we can obtain the equivalent local microforce balance as
\begin{equation}
\xi^{\chi}_{,i}+\zeta^{\chi}+\zeta^{\chi \text{ex}}=0 . \label{eq-balance-mircoforce}
\end{equation}

Specifically, for the improper MT ($\chi=\eta_I$) and proper MT ($\chi=\varepsilon_{kj}^0$), the microforces can be written as~\cite{Landis2008A} 
\begin{equation}
\begin{cases}
\xi^{\chi} =\xi^{\eta_I}_i  & ~\\
\zeta^{\chi}= \zeta^{\eta_I} & \text{for} ~ \chi=\eta_I \\
\zeta^{\chi \text{ex}}= \zeta^{\eta_I \text{ex}}&~
\end{cases}, \quad
\begin{cases}
\xi^{\chi} =\xi_{ijk}  & ~\\
\zeta^{\chi}= \zeta_{jk} & \text{for} ~ \chi=\varepsilon_{kj}^0 \\
\zeta^{\chi \text{ex}}= \zeta^\text{ex}_{jk} &~
\end{cases}.
\end{equation}

\item Balance of energy
  
The principle of the balance of energy can be stated as: the rate of change specific internal energy density $e$ is equal to the sum of power due to external forces and the heat input to the system. In solid with negligible inertia, the thermal heat flux $j_i^{\text{h}}$ and the heat source per unit volume $q^{\text{h}}$ are considered, and the internal microforce does not contribute to the energy change.

The first law of thermodynamics can be presented in the form of energy balance equation as
\begin{equation}
\int_\Omega \dot{e}  \text{d}V = \int_{\partial\Omega}(\sigma_{ji}\dot{u}_j+\xi^{\chi}\dot{\chi}-j_i^{\text{h}} )n_i \text{d}A + \int_\Omega(b_i\dot{u}_i +\zeta^{\chi \text{ex}}\dot{\chi}+q^{\text{h}} )\text{d}V, \label{eq-global-first}
\end{equation}
where $u_j$ is the displacement, and $n_i$ is the unit outer normal to $\partial\Omega$. 
During loading or unloading, the elastocaloric material will release or absorb the latent heat associated with the stress-induced MT. Herein, $q^{\text{h}}$ is the release/absorb rate of internal heat source from MT, which is only assumed to be related to the evolution rate ($\dot{\chi}$) of phase transformation~\cite{cisse2020asymmetric,nishiyama1958temperature}. 
For example, the $q^{\text{h}}$ for both $\eta_I$ and $\varepsilon_{kj}^0$ in cubic to tetragonal MT can be written as
\begin{equation}
q^{\text{h}} = 
\begin{cases}
Q\sum^I \dot{\eta_I} & \text{for} \quad \chi=\eta_I \\
Q^e(\dot{\varepsilon}_{11}^0 -\dot{\varepsilon}_{22}^0 ) & \text{for} \quad \chi=\varepsilon_{kj}^0
\end{cases},
\end{equation}
where $Q$ and $Q^e$ are the latent heat and play a critical role in the temperature change during MT.  
By Gauss law which converts the surface integration into volume one, Eq.~\ref{eq-global-first} can be rewritten as
\begin{equation}
\int_\Omega \dot{e}\text{d}V = \int_\Omega(\sigma_{ji}\dot{\varepsilon}_{ij}+\sigma_{ji,i}\dot{u}_j+\xi_{,i}^{\chi}\dot{\chi}+\xi^{\chi}\dot{\chi}_{,i}-j_{i,i}^\text{h}+b_i\dot{u}_i+\zeta^{\chi \text{ex}}\dot{\chi}+q^{\text{h}} ) \text{d}V,
\end{equation}
where $\dot{\varepsilon}_{ij}$ is the total strain rate with $\dot{\varepsilon}_{ij}=\frac{1}{2}(\dot{u}_{i,j}+\dot{u}_{j,i})$.
Then, combining Eqs.~\ref{eq-mecheq},~\ref{eq-angular-m} and~\ref{eq-balance-mircoforce}, as well as considering that the equation holds for any arbitrary volume, the energy balance equation in the local form reads
\begin{equation}
\dot{e}=\sigma_{ij}\dot{\varepsilon}_{ij}+\xi^{\chi}\dot{\chi}_{,i}-\zeta^{\chi}\dot{\chi}-j_{i,i}^{\text{h}}+q^{\text{h}} . \label{eq-local-first}
\end{equation}

\end{itemize}

\subsection{Constitutive relations and evolution equations} 
\begin{itemize}

\item Second law of thermodynamics

For the non-isothermal system, the second law of thermodynamics or entropy inequality, combining the global entropy balance with the Clausius--Duhem inequality~\cite{gurtin1966clausius} for the volume $\Omega$, is expressed as
\begin{equation}
\int_\Omega \dot{s} ~\text{d}V +\int_{\partial\Omega} \frac{j_i^{\text{h}}}{T}n_i \text{d}A- \int_\Omega \frac{q^{\text{h}}}{T} \text{d}V \geq 0 \label{eq-gener-second}
\end{equation}
where $s$ is the specific entropy per unit volume and $T$ is the temperature. Converting the surface integration in Eq.~\ref{eq-gener-second} into volume integration and considering its validity in any volume, we can obtain
\begin{equation}
\dot{s} +\Big(\frac{j_i^{\text{h}}}{T}\Big)_{,i}-\frac{q^{\text{h}}}{T} \geq 0 . \label{eq-gener-local-second}
\end{equation}

\item Free energy imbalance

Herein, the Helmholtz free energy is chosen as a proper thermodynamic potential in the non-isothermal system.
The Helmholtz free energy density $f$ per unit volume is defined by 
\begin{equation}
f=e-Ts \label{eq-f-e}
\end{equation}

Taking time derivatives at both sides, we get the relation
\begin{equation}
\frac{1}{T}\frac{\partial f}{\partial t}=\frac{1}{T}\frac{\partial e}{\partial t}-\frac{s}{T}\frac{\partial T}{\partial t}-\frac{\partial s}{\partial t}. \label{eq-time-derivate}
\end{equation}

Substituting the internal energy balance equation Eq.~\ref{eq-local-first} and the second law of thermodynamics Eq.~\ref{eq-gener-local-second} into the above relation Eq.~\ref{eq-time-derivate}, we can get an inequality
\begin{equation}
\frac{1}{T}\frac{\partial f}{\partial t} \leq \frac{1}{T}\sigma_{ij}\dot{\varepsilon}_{ij}+\frac{1}{T}\Big( \xi^{\chi}\dot{\chi}_{,i}-\zeta^{\chi}\dot{\chi} \Big)-\frac{s}{T}\frac{\partial T}{\partial t} + j_i^{\text{h}}\Big(\frac{1}{T}\Big)_{,i} \label{eq-energy-imbalance}
\end{equation}
the inequality in Eq.~\ref{eq-energy-imbalance} is referred as the free energy imbalance. It plays an analogous role to Eq.~\ref{eq-gener-local-second} in placing restrictions on the constitutive relations.

\item Coleman--Noll type analysis

In order to close the model, the constitutive relations for the Cauchy stress, the internal energy density, the entropy density, the heat flux, and the microforces should be provided. In this part, we derive the explicit form of the constitutive relations and the kinetic equations in terms of a thermodynamic potential. In this derivation, the Coleman--Noll argument~\cite{coleman1974thermodynamics} is applied so that the resulting constitutive relations will be thermodynamically consistent.
Invoking Truesdell's principle of equipresence~\cite{truesdell2004non}, it is reasonable to assume that $f$, $s$, $e$, $\sigma_{ij}$,  $j_i^{\text{h}}$, $\xi^{\chi}$ and $\zeta^{\chi}$ depend on $\varepsilon_{ij}$, $\chi$, $\dot{\chi}$, $\chi_{,i}$, $T$, $T_{,i}$. Specifically, the Helmholtz free energy density $f$ can be written as
\begin{equation}
f=f(\varepsilon_{ij}, \chi, \dot{\chi}, \chi_{,i}, T, T_{,i}) \label{eq-f-depend}
\end{equation}

The time derivative of $f$ and the chain rule lead to
\begin{equation}
\frac{\partial f}{\partial t}= \frac{\partial f}{\partial \varepsilon_{ij}}\frac{\partial \varepsilon_{ij}}{\partial t}+ \frac{\partial f}{\partial \chi}\frac{\partial \chi}{\partial t}+  \frac{\partial f}{\partial \dot{\chi}}\frac{\partial \dot{\chi}}{\partial t} + \frac{\partial f}{\partial \chi_{,i}}\frac{\partial \chi_{,i}}{\partial t}+ \frac{\partial f}{\partial T}\frac{\partial T}{\partial t}+ \frac{\partial f}{\partial T_{,i}}\frac{\partial T_{,i}}{\partial t} . \label{eq-chain-eule}
\end{equation}

Now substituting Eq.~\ref{eq-chain-eule} into the free energy imbalance Eq.~\ref{eq-energy-imbalance}, and grouping terms together, we can get
\begin{equation}
\frac{1}{T}\Big(\frac{\partial f}{\partial \varepsilon_{ij}}-\sigma_{ij}  \Big)\dot{\varepsilon}_{ij}+ \frac{1}{T}\Big(\frac{\partial f}{\partial \chi}+  \zeta^{\chi}\Big) \dot{\chi}+ \frac{1}{T} \frac{\partial f}{\partial \dot{\chi}}\frac{\partial \dot{\chi}}{\partial t} + \frac{1}{T}\Big(\frac{\partial f}{\partial \chi_{,i}}-\xi^{\chi} \Big)\dot{\chi}_{,i}+ \frac{1}{T}\Big(\frac{\partial f}{\partial T}+s\Big) \dot{T} + \frac{1}{T}\frac{\partial f}{\partial T_{,i}} \dot{T}_{,i}-j_i^{\text{h}} \Big(\frac{1}{T}\Big)_{,i} \leq 0 \label{eq-last-second}
\end{equation}

Here we provide an analysis of Eq.~\ref{eq-last-second} by invoking the arguments made by Coleman and Noll~\cite{coleman1974thermodynamics}, and notice that Eq.~\ref{eq-last-second} is linear with respect to $\dot{\varepsilon}_{ij}$, $\partial\dot{\chi}/\partial t$, $\dot{\chi}_{,i}$, $\dot{T}$, $\dot{T}_{,i}$. 
Hence, to satisfy Eq.~\ref{eq-last-second} in any admissible thermodynamics process, the coefficients of linear terms must vanish and thus these constitutive relations can be obtained
\begin{equation}
\sigma_{ij}=\frac{\partial f}{\partial \varepsilon_{ij}}, \quad \xi^{\chi}=\frac{\partial f}{\partial \chi_{,i}}, \quad \frac{\partial f}{\partial \dot{\chi}}=0, \quad s=-\frac{\partial f}{\partial T}, \quad \frac{\partial f}{\partial T_{,i}}=0. \label{eq-constitutive}
\end{equation}

With the relations in Eq.~\ref{eq-constitutive}, $f$ is independent of $\dot{\chi}$ and $T_{,i}$, and so Eq.~\ref{eq-f-depend} can be rewritten as
$f=f(\varepsilon_{ij}, \chi, \chi_{,i}, T)$.
Then, the left nonlinear terms in Eq.~\ref{eq-last-second} are reduced to
\begin{equation}
\frac{1}{T} \Big(\frac{\partial f}{\partial \chi}+\zeta^{\chi} \Big)\dot{\chi}-j_i^{\text{h}}\Big(\frac{1}{T}\Big)_{,i} \leq 0 . \label{eq-reducesecond-21}
\end{equation}

\item Evolution equations

The inequality in Eq.~\ref{eq-reducesecond-21} can be satisfied by assuming the following relations
\begin{equation}
\dot{\chi} = -L \left( \frac{\partial f}{\partial \chi}+\zeta^{\chi} \right) \label{eq-common-eta}
\end{equation}
\begin{equation}
j_i^{\text{h}}=\kappa_{ij}\left(\frac{1}{T}\right)_{,j}
\end{equation}
where positive $L$ and positive semi-definite $\kappa_{ij}$ are coefficients with respect to temperature. 
Further, combining Eqs.~\ref{eq-balance-mircoforce},~\ref{eq-constitutive} with~\ref{eq-common-eta}, the evolution equations for non-conserved OPs $\chi$ can be written as
\begin{equation}
\dot{\chi}= -L \left[\frac{\partial f}{\partial \chi}- \left(\frac{\partial f}{\partial \chi_{,i}} \right)_{,i} - \zeta^{\chi\text{ex}} \right]. \label{eq-AC}
\end{equation}

The Eq.~\ref{eq-AC} coincides with the general Allen--Cahn equation~\cite{allen1979microscopic}. In addition, the heat flux equation reads as
\begin{equation}
j_i^{\text{h}}=-K_{ij} T_{,j} \label{eq-heat-flux}
\end{equation}
where $K_{ij}=\kappa_{ij}/T^2$ is a tensor representing the thermal conductivity.



\end{itemize}

\subsection{Helmholtz free energy}

In the non-isothermal PFM, Helmholtz free energy density $f$ of the system with microstructure evolution consists of the chemical free energy  density $f^{\text{chem1}}+f^{\text{chem2}}$, the gradient free energy  density $f^{\text{grad}}$, and the elastic energy  density $f^{\text{ela}}$, i.e.,
\begin{equation}
f(\varepsilon_{ij},\chi,\chi_{,i},T) = f^{\text{chem1}}+f^{\text{chem2}}+f^{\text{grad}}+f^{\text{ela}}.
\end{equation} \label{eq2}

\subsubsection{Improper MT}
The chemical free energy or Landau free energy $f^{\text{chem1}}$ of a closed system containing austenite and martensite is determined by the distribution of OPs $\eta_I$.
The minimum energy states at $\eta_I = 0$ and $\eta = 1$ represent the austenite phase and the corresponding martensite variant $I$, respectively.
The associated chemical energy can be constructed as a Landau polynomial~\cite{ohmer2022phase}, i.e.,
\begin{equation}
f^{\text{chem1}}=A(T) \sum^I\eta_I^2-B(T)\sum^I\eta_I^3+C(T) \Big(\sum^I\eta_I^2 \Big)^2
\end{equation}
where $A(T)$, $B(T)$, and $C(T)$ are temperature-dependent coefficients.

The gradient energy is often constructed to account for the interface between different phases, which can be expressed in terms of OPs' gradient, i.e.,
\begin{equation}
f^{\text{grad}} = \frac{1}{2}\beta^{\eta}(T)\sum^I \eta_{I,i} \eta_{I,i}
\end{equation}
where $\beta^{\eta}(T)$ is the gradient energy coefficient depending on temperature.

The total strain tensor $\varepsilon_{ij}$ is given as
\begin{equation}
\varepsilon_{ij}=\varepsilon_{ij}^{\text{ela}}+\varepsilon_{ij}^{\text{tr}} + \varepsilon_{ij}^{\text{th}}
\end{equation} 
where $\varepsilon_{ij}^{\text{ela}}$ is the elastic strain, $\varepsilon_{ij}^{\text{tr}}$ is the transformation strain generated by structural transformation, and $\varepsilon_{ij}^{\text{th}}$ is the thermal strain caused by temperature change. Therefore, the elastic energy can be expressed as
\begin{equation}
f^\text{ela} =\frac{1}{2} C_{ijkl}\varepsilon_{ij}^{\text{ela}}\varepsilon_{kl}^{\text{ela}}= \frac{1}{2} C_{ijkl} (\varepsilon_{ij} - \varepsilon_{ij}^{\text{tr}}-\varepsilon_{ij}^{\text{th}}) (\varepsilon_{kl} - \varepsilon_{kl}^{\text{tr}}-\varepsilon_{kl}^{\text{th}}) 
\end{equation}
in which $C_{ijkl}=\sum^I \Phi(\eta_I) C_{ijkl}^{\eta_I}+\big(1-\sum^I\Phi(\eta_I)\big)C_{ijkl}^A$ with the interpolation function $\Phi(\eta_I)=\eta_I^3(10-15\eta_I+6\eta_I^2)$.
$C_{ijkl}^{\eta_I}$ and $C_{ijkl}^A$ are the elastic tensors of the $I^{\text{th}}$ martensitic variant and austenite, respectively.

The chemical energy $f^{\text{chem2}}$ represents the main concave term of the free energy and is related to the heat conduction (e.g., the internal energy density from temperature change). 
The complete and general form of the $f^{\text{chem2}}$ can be formulated as
\begin{equation}
f^{\text{chem2}}=-c_1 T\text{ln}(T+1)- \frac{1}{2} c_2T^2-\frac{1}{6}c_3T^3 + \cdots \label{eq-fchem2-improper}
\end{equation}
where $c_i=\sum^I \Phi(\eta_I) c_i^{\eta_I}+(1-\sum^I\Phi(\eta_I))c_i^A$ with $c_i^{\eta_I}$ and $c_i^A$ as the $i^\text{th}$ coefficients for the specific heat (Eq.~\ref{eqcv}) of $I^\text{th}$ martensitic variant and austenite, respectively.
Meanwhile, the specific heat per unit volume is expressed as a polynomial function of temperature, i.e.,
\begin{equation}
c_\text{v}  = c_1 + c_2 T + c_3 T^2 + \cdots .  \label{eqcv}
\end{equation}



In addition, in order to derive the general kinetic equation for temperature $T$, combining
Eqs.~\ref{eq-balance-mircoforce},~\ref{eq-local-first},~\ref{eq-constitutive} and~\ref{eq-heat-flux}, the evolution equation of $e$ reads as
\begin{equation}
\dot{e}= \sigma_{ij}\dot{\varepsilon}_{ij}+ \left(\frac{\partial f}{\partial \eta_{I,i}}\dot{\eta_I}\right)_{,i}+\zeta^{\eta_I \text{ex}}\dot{\eta_I}+ \left(K_{ij}T_{,j}\right)_{,i} +q^{\text{h}}. \label{eq-e-evolution}
\end{equation}

Neglecting the first three terms on the right side of Eq.~\ref{eq-e-evolution}, one yields the common governing equation for temperature as
\begin{equation}
c_v\dot{T}= \big(K_{ij}T_{,j}\big)_{,i} +q^{\text{h}}.
\end{equation}

\subsubsection{Proper MT}

As an example here, the OPs of eigen strain $\varepsilon_{ij}^{0}$ is considered to model the proper MT.
The corresponding chemical free energy density $f^{\text{chem1}}$ describing the proper MT can be represented as~\cite{Zhang2005Phase}
\begin{equation}
f^{\text{chem1}}= Q_1e_1^2+Q_2(e_2^2+e_3^2)-Q_3(e_3^3-3e_3e_2^2)+Q_4(e_2^2 + e_3^2)^2+Q_5(e_4^2+e_5^2+e_6^2), \label{eq-chem-eps}
\end{equation}
where $Q_1$ and $Q_5$ are bulk and shear modulus, respectively. The coefficients $Q_2$, $Q_3$, and $Q_4$ are Landau constants determining the transition temperature $T_0$ and the transformation strains in the product phase.
$e_i$ are the symmetry-adapted strain defined in term of the transformation strains as
\begin{equation}
\begin{split}
& e_1=(\varepsilon_{11}^0+\varepsilon_{22}^0+\varepsilon_{33}^0)/\sqrt{3},   \quad ~ e_4=\varepsilon_{12}^0, \\
& e_2=(\varepsilon_{11}^0-\varepsilon_{22}^0)/\sqrt{2},   \quad\quad\quad~~ e_5=\varepsilon_{23}^0, \\
& e_3=(2\varepsilon_{33}^0-\varepsilon_{22}^0-\varepsilon_{11}^0)/\sqrt{6},   \quad e_6=\varepsilon_{13}^0. \\
\end{split}
\end{equation}
In addition, the $f^{\text{chem2}}$ is similar to the case of the improper MT in Eq.~\ref{eq-fchem2-improper}.

The austenite and martensite can be described by the eigen strain $\varepsilon_{ij}^0$. For instance, $\varepsilon_{23}^0=\varepsilon_{13}^0=\varepsilon_{12}^0=0$ is set for a cubic to tetragonal MT. Further, the austenite (cubic) is represented as ($\varepsilon_{11}^0=0$,~$\varepsilon_{22}^0=0$,~$\varepsilon_{33}^0=0$), the three tetragonal martensitic variants are described by tet$_1$~=~$(-\varepsilon_0,~1/2\varepsilon_0,~1/2\varepsilon_0)$, tet$_2$~=~$(1/2\varepsilon_0,~-\varepsilon_0,~1/2\varepsilon_0)$, and tet$_3$~=~$(1/2\varepsilon_0,~1/2\varepsilon_0,~-\varepsilon_0)$, where $\varepsilon_0$ is the magnitude of the spontaneous strain at a given temperature.

Similarly, the gradient energy density can be written as
\begin{equation}
f^{\text{grad}} = \frac{1}{2} \beta^e(T) \sum_{i=1}^3\sum_{j=1}^3\varepsilon_{ii,j}^0\varepsilon_{ii,j}^0,
\end{equation}
where $\beta^e(T)$ is the strain gradient coefficient. The elastic strain energy density can be written as
\begin{equation}
f^{\text{ela}} = \frac{1}{2} C_{ijkl}(\varepsilon_{ij}-\varepsilon_{ij}^0-\varepsilon_{ij}^{\text{th}})(\varepsilon_{kl}-\varepsilon_{kl}^0-\varepsilon_{kl}^{\text{th}}).
\end{equation}


\section{Application of non-isothermal PFM to Mn-22Cu alloy} \label{sec-application}  
In this section, we apply the above non-isothermal phase-field framework to a model Mn-22Cu alloy and present the detailed formulations for the total free energy, constitutive relations, and governing or evolution equations. In Mn-22Cu SMA, there exist a face-centered cubic high-symmetry austenitic phase at high temperature and three variants of a face-center tetragonal low-symmetry martensitic phase at low temperature~\cite{daly2007stress,shimizu1982crystallographic}. 

\subsection{Improper MT}

The three martensitic variants are energetically equivalent. Herein we choose non-conserved OPs $\eta_I$ ($I = 1, 2, 3$) to represent the improper MT in PFM, and the value of $\eta_I$ varies from 0 to 1.

\subsubsection{Chemical free energy density}
\begin{figure}[!t]
\centering
  \includegraphics[width=16cm]{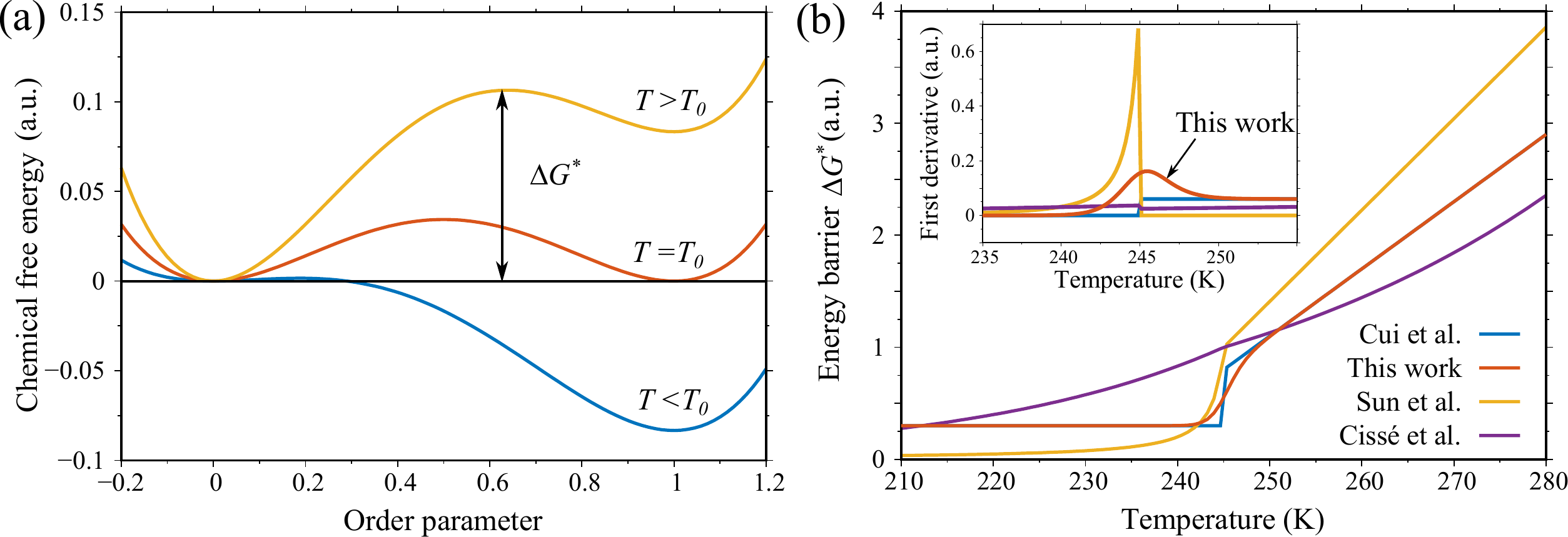}
  \caption{(a) Chemical free energy density as a function of order parameter. (b) Energy barrier functions from literature~\cite{cui2017three,sun2019non,cisse2020elastocaloric} compared with what we propose in this work.} \label{f-energy}
\end{figure}

The chemical free energy $f^{\text{chem1}}$ represents the chemical driving force of the MT in a stress-free Mn-22Cu alloy, which can be expressed as a Landau 2-3-4 polynomial
\begin{equation}
\begin{split}
f^\text{chem1} =  A(T)(\eta_1^2+\eta_2^2+\eta_3^2) - B(T) (\eta_1^3+\eta_2^3+\eta_3^3) +C(T) (\eta_1^2+\eta_2^2+\eta_3^2)^2 
\end{split}
\end{equation}
where $A(T)$, $B(T)$ and $C(T)$ are positive temperature-dependent coefficients, expressed as $A(T)=16\Delta G^*$, $B(T)=A(T)-4\Delta G_{\text{m}}$ and $C(T)=0.5A(T)-3\Delta G_{\text{m}}$. $\Delta G^*$ is the temperature-dependent energy barrier between austenite and martensite. $\Delta G_{\text{m}}$ is the driving force of MT. The temperature-dependent profile of $f^{\text{chem1}}$ in non-isothermal system and the different models for austenite-martensite energy barriers are shown in Fig.~\ref{f-energy}a and b, respectively. $\eta_I=0$ or 1 represents the system stable or metastable from the principle of minimization of the free energy. More explicitly, OPs $\eta_I=1$ indicates the $I^{\text{th}}$ martensitic variant and $\eta_I=0$ ($I=1,2,3$) represents austenite.

Since the MT is a first-order diffusionless structural transformation and is not sharp in real materials~\cite{manosa2017materials}, there are several ways in the literature~\cite{cisse2020elastocaloric,cui2017three,sun2019non} to formulate the energy barrier function $\Delta G^*$. For instance, Cui et al.~\cite{cui2017three} proposed that
\begin{equation}
\Delta G^*=
\begin{cases}
0.3Q / 32& T\leq T_0\\
[0.8+0.06(T-T_0)]Q / 32& T>T_0
\end{cases} ,
\end{equation}
Sun et al.~\cite{sun2019non} proposed that
\begin{equation}
\Delta G^*=
\begin{cases}
Q/\big[32\big(k_1-k_2(T-T_0)\big)\big]& T\leq T_0\\
k_3Q(T-T_0)/T_0+Q/32k_1& T>T_0
\end{cases} ,
\end{equation}
and Ciss{\'e} et al.~\cite{cisse2020elastocaloric} proposed that
\begin{equation}
\Delta G^*=
\begin{cases}
Q \text{exp}\big[a_1(T-T_0)/T_0 \big]& T\leq T_0\\
Q \text{exp}\big[a_2(T-T_0)/T_0 \big]& T>T_0
\end{cases} 
\end{equation}
where $Q$ is the specific latent heat and $T_0$ is the chemical equilibrium temperature.
It can be seen from Fig.~\ref{f-energy}b that the first derivative of all these $\Delta G^*$ functions~\cite{cisse2020elastocaloric,cui2017three,sun2019non} are discontinuous at $T_0$, leading to the problematic calculation of driving force and eCE at $T_0$. Herein, we formulate the energy barrier as a smooth function of temperature. Specifically, we take advantage of the hyperbolic tangent function to modify the piecewise $\Delta G^*$ function, i.e.,

\begin{equation}
\Delta G^*=  \frac{0.3Q}{64}\left[ 1-\text{tanh} \left( \frac{T-T_0}{\delta T} \right) \right]  +\frac{\big[ 0.8+0.06(T-T_0) \big] Q}{64}\left[ 1+\text{tanh} \left( \frac{T-T_0}{\delta T} \right) \right].\label{eq4}
\end{equation}

$\Delta G_{\text{m}}$ is also a continuous function of temperature, i.e.,
\begin{equation}
\Delta G_{\text{m}} = \frac{Q(T-T_0)}{T_0}.
\end{equation}

It should be noted that $\delta T$ is a new parameter associating with the energy barrier, which could be adjusted according to the transformation temperature window from experimental results. In this work, we assume a moderately sharp transition and set $\delta T = 2$~K.
As shown in Fig.~\ref{f-energy}b, compared to the functions proposed in literature~\cite{cisse2020elastocaloric,cui2017three,sun2019non}, our modified energy barrier function is continuous and differentiate at $T_0$. This modification could resolve the difficulty of calculating eCE by indirect and direct method at the transition point.

In addition, the internal energy density from temperature change $f^{\text{chem2}}$ reads
\begin{equation}
f^{\text{chem2}}= c_\text{v} T \text{ln}(T+1) \label{eq-f-th-classic}
\end{equation}
where $c_\text{v}=\sum_{I=1}^3 \Phi(\eta_I) c_\text{v}^{\eta_I}+(1-\sum_{I=1}^3\Phi(\eta_I))c_\text{v}^A$ is the specific heat per unit volume. $c_\text{v}^{\eta_I}$ and $c_\text{v}^A$ are the specific heat per unit volume of the $I^{\text{th}}$ martensitic variant and austenite, respectively. The difference between $c_\text{v}^{\eta_I}$ and $c_\text{v}^A$ is insignificant for Mn-22Cu alloy so that they approximately take the same values.

\subsubsection{Gradient energy density}
The gradient energy density can be expressed as a function of the OPs' gradient~\cite{artemev2001three}, i.e.,
\begin{equation}
f^\text{grad} = \frac{1}{2} \beta^{\eta}(T) \sum_{I=1}^3 \eta_{I,i}\eta_{I,i} \label{eq6}
\end{equation}
where $\beta^{\eta}(T)$ is the gradient energy coefficient related to the interfacial energy and interface thickness~\cite{yeddu2012three-dimensional}. 

\subsubsection{Elastic strain energy  density}

The elastic strain energy  density in Mn-22Cu alloys can be given as
\begin{equation}
f^\text{ela} = \frac{1}{2} C_{ijkl} (\varepsilon_{ij} - \varepsilon_{ij}^{\text{tr}}-\varepsilon_{ij}^{\text{th}}) (\varepsilon_{kl} - \varepsilon_{kl}^{\text{tr}}-\varepsilon_{kl}^{\text{th}})  
\end{equation}
where $C_{ijkl}$ is the component of the fourth-order elastic tensor. In this work, $c_{ij}$ in Table~\ref{t1} denotes the Voigt notation of $C_{ijkl}$, i.e., $c_{11}=C_{1111}, c_{12}=C_{1122}, c_{44}=C_{1212}$.

We incorporate $\varepsilon_{ij}^{\text{tr}}$ and $\varepsilon_{ij}^{\text{th}}$ into the PFM. 
$\varepsilon^{\text{tr}}_{ij}$ can be expressed as the stress-free eigen strain, which is in general defined as
\begin{equation}
\varepsilon_{ij}^\text{tr} = \varepsilon_{ij}^{00}(1)\eta_1 + \varepsilon_{ij}^{00}(2)\eta_2 + \varepsilon_{ij}^{00}(3)\eta_3.
\end{equation}
$\varepsilon^{00}(i)$ ($i=1, 2, 3$) is determined by the orientation relationship and lattice distortion between martensite and austenite with regard to the FCC-FCT MT. It can be simplified to~\cite{artemev2001three}
\begin{equation}
\begin{split}
 \left[ \varepsilon_{ij}^{00}(1) \right] =\left[
\begin{array}{ccc}
\varepsilon_3 & 0 & 0 \\
0 & \varepsilon_1 & 0 \\
0 & 0 & \varepsilon_1
\end{array}
\right] ,\quad
\left[ \varepsilon_{ij}^{00}(2) \right] =\left[
\begin{array}{ccc}
\varepsilon_1 & 0 & 0 \\
0 & \varepsilon_3 & 0 \\
0 & 0 & \varepsilon_1
\end{array}
\right]  ,\quad
\left[ \varepsilon_{ij}^{00}(3) \right] =\left[
\begin{array}{ccc}
\varepsilon_1 & 0 & 0 \\
0 & \varepsilon_1 & 0 \\
0 & 0 & \varepsilon_3
\end{array}
\right]
\end{split}
\end{equation}
where $\varepsilon_1=(a-a_c)/a_c $ and $\varepsilon_3=(c-a_c)/a_c$. We choose $\varepsilon_1=0.01$ and $\varepsilon_3=-0.02$~\cite{cui2017three} in this work. 

In addition, the thermal strain is computed as
\begin{equation}
\varepsilon_{ij}^{\text{th}}=\alpha_{ij} (T-T^{\text{ref}})
\end{equation}
where $\alpha_{ij}=\sum_{I=1}^3 \Phi(\eta_I) \alpha^{\eta_I}\delta_{ij} +\big(1-\sum_{I=1}^3\Phi(\eta_I) \big)\alpha^A\delta_{ij}$ is a tensor representing thermal expansion, in which $\alpha^{\eta_I}$ and $\alpha^A$ are thermal expansion coefficients for the $I^{\text{th}}$ martensitic variant and austenite, respectively. $T^{\text{ref}}$ is the reference temperature at which there is zero thermal strain. In the PF simulations, $T^{\text{ref}}$ is the initial temperature.

\subsubsection{Constitutive relations for PFM  of Mn-22Cu alloy} \label{subsec-constitutive}

Using Eq.~\ref{eq-constitutive}, the constitutive relations for PFM of Mn-22Cu alloy is given by
\begin{equation}
\begin{split}
& \sigma_{ij}=\frac{\partial f}{\partial \varepsilon_{ij}}=C_{ijkl}(\varepsilon_{kl}-\varepsilon_{kl}^{\text{tr}} -\varepsilon_{kl}^{\text{th}}) \\
& \zeta^{\eta_I}_i = \frac{\partial f}{\partial \eta_{I,i}}= \beta^{\eta}(T)\eta_{I,i} \\
& e = f+ Ts =f-T \frac{\partial f}{\partial T} \\
& j_i^{\text{h}} =-K_{ij}T_{,j} 
\end{split}  \label{eq-constitutive-improper}
\end{equation}

The internal energy density $e$ for Mn-22Cu alloy can be simplified as
\begin{equation}
\begin{split}
e &=c_\text{v}\frac{T^2}{T+1}+ \left(A(T)-T\frac{\partial A(T)}{\partial T}\right)\sum_{I=1}^3\eta_I^2- \left(B(T)-T\frac{\partial B(T)}{\partial T}\right)\sum_{I=1}^3\eta_I^3+\left(C(T)-T\frac{\partial C(T)}{\partial T}\right)\Big(\sum_{I=1}^3\eta_I^2 \Big)^2 \\
& \quad + \frac{1}{2}C_{ijkl}\left(\varepsilon_{ij}^{\text{ela}}-T\frac{\partial \varepsilon_{ij}^{\text{ela}}}{\partial T} \right)\varepsilon_{kl}^{\text{ela}} +\frac{1}{2}\left(\beta^{\eta}(T)-T\frac{\partial \beta^{\eta}(T)}{\partial T}\right)\sum_{I=1}^3 \eta_{I,i}\eta_{I,i} ~. \\
\end{split} \label{eq-e-Mn}
\end{equation}

\subsubsection{Governing equations for PFM  of Mn-22Cu alloy} \label{sec-governing}

We adopt Eq.~\ref{eq-AC} to govern the spatial and temporal evolution of $\eta_I$, which is analogous to the time-dependent Ginzburg-Landau (TDGL) kinetic equation~\cite{de1993crossover,man2010microstructural}. Further, using Eqs.~\ref{eq-e-evolution} and \ref{eq-e-Mn}, and assuming quasi-static mechanics, the temperature evolution equation in non-isothermal PFM can be obtained. Finally, The governing equations are deduced and summarized as
\begin{equation}
\begin{split}
& \sigma_{ij,j}=0     \\
&\dot{\eta_I} =-L^{\eta}\left( \frac{\partial f^{\text{chem}}}{\partial\eta_I}+\frac{\partial f^{\text{ela}}}{\partial\eta_I} - \beta^{\eta} \eta_{I,ii} - \zeta^{\eta_I\text{ex}} \right)     \\
&\dot{e} = \left(\beta^{\eta}\sum_{I=1}^3\eta_{I,i} \dot{\eta_I}\right)_{,i}+\zeta^{\eta_I \text{ex}}\dot{\eta_I}+ K_{ij,i}T_{,j}+K_{ij}T_{,ij} + Q\sum_{I=1}^3 \dot{\eta_I} .
\end{split} \label{eq-governing-equations}
\end{equation}
where $L^{\eta}$ is the kinetic coefficient characterizing the interfacial migration. 


\subsection{Proper MT}
For the proper MT, the OPs $\varepsilon_{ii}^0$ are used to directly describe the austenite and martensite. The corresponding chemical free energy or Landau free energy density for 2D domain in cubic-tetragonal MT of Mn-22Cu can be represented as~\cite{Dhote2012Dynamic}
\begin{equation}
f^{\text{chem1}}= \frac{1}{2}Q_1e_1^2+\frac{1}{2}Q_2(T)e_2^2+\frac{1}{2}Q_3e_3^2-\frac{1}{4}Q_4(T)e_2^4 + \frac{1}{6}Q_5(T)e_2^6, \label{eq-chem-eps-1}
\end{equation}
where $e_1=(\varepsilon_{11}^0+\varepsilon_{22}^0)/\sqrt{2}$, $e_2=(\varepsilon_{11}^0-\varepsilon_{22}^0)/\sqrt{2}$, and $e_3=(\varepsilon_{12}^0+\varepsilon_{21}^0)/2$. These coefficients ($Q_1$ to $Q_5$) in Landau free energy could be obtained by fitting the experimental results. For cubic to tetragonal MT, the eigen strains for martensitic variants are
\begin{equation}
\varepsilon^{00}=\left(
\begin{array}{cc}
   -\varepsilon_0 &         0 \\
         0&    \varepsilon_0  
\end{array}
\right) ~\text{for V1, or}~
\left(
\begin{array}{cc}
   \varepsilon_0 &         0 \\
         0&    -\varepsilon_0  
\end{array}
\right)~\text{for V2},
\end{equation}
where $\varepsilon_0=0.02$ is the transformation strain in Mn-22Cu. Here, ($e_1=0, ~e_2=0, ~e_3=0$) represents for austenite, and ($e_1=0, ~e_2=0.04,~ e_3=0$) represents for martensitic variant 1 (V1), and ($e_1=0, ~e_2=-0.04,~ e_3=0$) represents for martensitic variant 2 (V2). For a first-order transition, the coefficients of $Q_4$ and $Q_5$ is positive and is needed for stability~\cite{jacobs2003simulations}. Here, we assume that the value of $Q_2(T),Q_4(T),Q_5(T)$ are similar to $A(T),B(T),C(T)$ in this work owing to the lack of experimental value for Mn-22Cu.
Thus, $Q_2(T),Q_4(T),Q_5(T)$ are assumed as
\begin{equation}
\begin{split}
&Q_2(T)=A(T)/(2 \varepsilon_0)^2 \\
&Q_4(T)=B(T)/(2 \varepsilon_0)^4 \\
&Q_5(T)=C(T)/(2 \varepsilon_0)^6 .
\end{split}
\end{equation}

Similar to the case of the improper MT, the internal energy density, gradient energy density, and elastic strain energy density can be written as
\begin{equation}
\begin{split}
&f^{\text{chem2}} = c_\text{v} T \text{ln}(T+1), \\
&f^{\text{grad}} = \frac{1}{2} \beta^e \left[ \left( \varepsilon_{11,1}^0 \right)^2 +\left( \varepsilon_{11,2}^0 \right)^2 +\left( \varepsilon_{22,1}^0 \right)^2 +\left( \varepsilon_{22,2}^0 \right)^2  \right], \\
&f^{\text{ela}} = \frac{1}{2} C_{ijkl}(\varepsilon_{ij}-\varepsilon_{ij}^0-\varepsilon_{ij}^{\text{th}})(\varepsilon_{kl}-\varepsilon_{kl}^0-\varepsilon_{kl}^{\text{th}}),
\end{split}
\end{equation}
respectively.

The constitutive relations of proper MT PFM are analogous to Eq.~\ref{eq-constitutive-improper}, and the governing equations can be written as
\begin{equation}
\begin{split}
& \sigma_{ij,j}=0     \\
& \dot{\varepsilon}_{ii}^0=-L^e \left(\frac{\partial f^{\text{chem}}}{\partial \varepsilon_{ii}^0} +\frac{\partial f^{\text{ela}}}{\partial\varepsilon_{ii}^0} - \beta^{e} \varepsilon_{ii,jj} - \zeta_{ii}^{\text{ex}}  \right)  \\
& \dot{e} = \left(\beta^{e}\sum_{i=1}^2\varepsilon_{ii,j} \dot{\varepsilon}_{ii}^0 \right)_{,j} + \zeta_{ii}^{\text{ex}} \dot{\varepsilon}_{ii}^0+ K_{ij,i}T_{,j}+K_{ij}T_{,ij} + Q^e(\dot{\varepsilon}_{11}^0-\dot{\varepsilon}_{22}^0). \label{eq-goverming-improper}
\end{split}
\end{equation}

\section{Finite-element implementation}\label{sec3}

Herein, we use finite element method to solve the governing equations in Eqs.~\ref{eq-governing-equations} and~\ref{eq-goverming-improper}, and convert the strong forms into weak forms by introducing a test function. Note that the degrees of freedom $\eta_I$ and $\varepsilon_{ii}^0$ are replaced by the generalized degree of freedom $\chi$ in the finite-element implementation.
Therefore, the degrees of freedom are set as $u_1,u_2,u_3,\chi,T$.
Assuming $K_{ij}$ is a constant $K$ in Eqs.~\ref{eq-governing-equations} and~\ref{eq-goverming-improper}, the weak forms are formulated as
\begin{equation}
\begin{split}
&0 = \int_\Omega \sigma_{ij}\phi_{i,j} \text{d}v - \int_{\partial\Omega}\sigma_{ij} n_j \phi_i \text{d}s \\
&0 = \int_\Omega \Big[ \psi \Big(\frac{\dot{\chi}}{L}+ \frac{\partial f^{\text{chem}}}{\partial\chi}+\frac{\partial f^{\text{ela}}}{\partial\chi}- \zeta^{\chi\text{ex}} \Big) + \psi_{,i} \beta \chi_{,i} \Big] \text{d}v - \int_{\partial\Omega} \psi \beta \chi_{,i}n_i  \text{d}s \\
&0 = \int_\Omega \Big[ \vartheta\Big(\dot{e}- q^{\text{h}} -\zeta^{\chi \text{ex}} \dot{\chi} \Big) + \vartheta_{,i}\Big(\beta\sum \chi_{,i}\dot{\chi} +K T_{,i} \Big) \Big] \text{d}v -\int_{\partial\Omega} \vartheta \Big( \beta\sum \chi_{,i}\dot{\chi}n_i + K T_{,i}n_i \Big) \text{d}s ,
\end{split} \label{eq-weak-forms}
\end{equation}
where $\phi_i$, $\psi$ and $\vartheta$ are the test function for $u_i$, $\chi$ and $T$, respectively.
Note that the surface terms ($\sigma_{ij}n_j$ and $T_{,i}n_i$) in Eq.~\ref{eq-weak-forms} represent the surface traction and heat flux boundary conditions. $n_i$ is the normal vector of the boundary $\partial\Omega$.

By introducing the shape functions for independent variables and test functions, the discretized equations can be written as 
\begin{equation}
\begin{split}
& u_i = N^L u_i^L \quad\quad \chi = N^L \chi^L \quad\quad \dot{\chi}=N^L \dot{\chi}_I^L \quad\quad T=N^L T^L \\
&\dot{T}=N^L \dot{T}^L \quad\quad \phi_i = N^L \phi_i^L  \quad\quad \psi = N^L \psi^L \quad\quad \vartheta = N^L \vartheta^L   \label{eq-57}
\end{split}
\end{equation}
where $L$ denotes the node number. $N^L$ is the shape function. Here we assume quasi-static mechanics, and neglect $\zeta^{\chi\text{ex}}$, and do not consider the dynamic PFM. After the insertion of Eq.~\ref{eq-57} into Eq.~\ref{eq-weak-forms}, the following elemental residuals can be obtained
\begin{equation}
\begin{split}
& R_{u_i}^L = \int_\Omega \sigma_{ij} N_{,j}^L \text{d}v - \int_{\partial\Omega} N^L\sigma_{ij}n_j \text{d}s  \\
& R_{\chi}^L=\int_\Omega \left[N^L\left( \frac{1}L\dot{\chi}+\frac{\partial f^\text{ela}}{\partial \chi} +\frac{\partial f^\text{chem}}{\partial \chi} \right) + \beta \chi_{,i} N_{,i}^L \right]\text{d}v - \int_{\partial \Omega} N^L \beta \chi_{,i}n_i \text{d}s  \\
& R_T^L=\int_\Omega\Big[N^L\Big(\dot{e}-q^{\text{h}}\Big)+N_{,i}^L\Big(\beta\sum \chi_{,i}\dot{\chi} +K T_{,i} \Big)  \Big] \text{d}v - \int_{\partial\Omega} N^L \Big( \beta\sum \chi_{,i}\dot{\chi}n_i + K T_{,i}n_i \Big) \text{d}s.
\end{split}
\end{equation}\label{eq23}

With regard to the time dependence of the residuals, we use the implicit backward Euler method to realize the time discretization~\cite{yi2014constraint}. The residual equation for the current time step $t_{n+1}$ is
\begin{equation}
\underline{\mathbf{R}}_{n+1}^L=\underline{\mathbf{R}}^L \left(\underline{\mathbf{d}}_{n+1}^J~,\frac{\underline{\mathbf{d}}_{n+1}^J-\underline{\mathbf{d}}_n^J}{\Delta t} \right),
\end{equation}
where $(\underline{\mathbf{d}}_{n+1}^J-\underline{\mathbf{d}}_n^J)/\Delta t = \underline{\mathbf{\dot{d}}}_{n+1}^J$ and $\Delta t$ is time step. $\underline{\mathbf{d}}_{n+1}^J$ should be solved in this equation.
For solving these non-linear equations, the Newton iteration scheme is performed at each time step. The corresponding iteration matrix is
\begin{equation}
\underline{\mathbf{S}}^{LJ}=\underline{\mathbf{K}}^{LJ}+\frac{1}{\Delta t}\underline{\mathbf{D}}^{LJ},
\end{equation}
where $\underline{\mathbf{K}}^{LJ}$ is the stiffness matrix and $\underline{\mathbf{D}}^{LJ}$ is the damping matrix. 
We adopt the open source Multiphysics Object Oriented Simulation Environment (MOOSE)~\cite{tonks2012object} to implement this PFM. 

In order to evaluate the eCE of polycrystalline material, we make use of the rotation matrix to establish the polycrystalline PFM. 
The rotation matrices spinning around the $x$-axis, $y$-axis and $z$-axis are given by
\begin{equation}
\begin{split}
&\underline{\mathbf{Q}}_x =
\left(             
\begin{array}{ccc}  
1& 0  & 0 \\
0& \text{cos}\alpha & \text{sin}\alpha \\
0 & -\text{sin}\alpha & \text{cos}\alpha \\
\end{array}
\right), \quad
 \underline{\mathbf{Q}}_y=
\left(             
\begin{array}{ccc}  
\text{cos}\beta & 0  & -\text{sin}\beta \\
0 & 1 & 0 \\
\text{sin}\beta  & 0 & \text{cos}\beta \\
\end{array}
\right), \quad
\underline{\mathbf{Q}}_z=
\left(             
\begin{array}{ccc}  
\text{cos}\gamma & \text{sin}\gamma  & 0 \\
-\text{sin}\gamma & \text{cos}\gamma & 0 \\
0 & 0 & 1 \\
\end{array}
\right) 
\end{split} \label{eq17}
\end{equation}
where $\alpha, \beta, \gamma$ are the Euler angles. The three-dimensional rotation matrix $\underline{\mathbf{Q}}$ is given by
\begin{equation}
\underline{\mathbf{Q}}=\underline{\mathbf{Q}}_z\underline{\mathbf{Q}}_y\underline{\mathbf{Q}}_x
\end{equation}

The relationship between the global and the local variables are
\begin{equation}
\begin{split}
& \underline{\mathbf{u}}= \underline{\mathbf{Q}}~\underline{\mathbf{u}}^\prime \\
& \underline{\boldsymbol{\sigma}} = \underline{\mathbf{K}}_{\sigma} \underline{\boldsymbol{\sigma}}^\prime  \\
& \underline{\boldsymbol{\varepsilon}} = \underline{\mathbf{K}}_{\varepsilon}~\underline{\boldsymbol{\varepsilon}}^\prime  \label{eq19}
\end{split}
\end{equation}
in which $u$, $\varepsilon$ and $\sigma$ are the values in the global coordinate, while $u^\prime$, $\varepsilon^\prime$ and $\sigma^\prime$ are values in the grain local coordinate. $\underline{\mathbf{K}}_{\sigma}$ and $\underline{\mathbf{K}}_{\varepsilon}$ are the stress rotation matrix and strain rotation matrix, respectively.
The eigen strain ($\varepsilon^{\text{tr}}_{ij}$) and elastic tensor ($C_{ijkl}$) should be rotated by $\underline{\mathbf{K}}_{\sigma}$ and $\underline{\mathbf{K}}_{\varepsilon}$. It is not difficult to deduce $\underline{\mathbf{K}}_{\varepsilon}$ and $\underline{\mathbf{C}}'$ from Eq.~\ref{eq19}  that
\begin{equation}
\begin{split}
& \underline{\mathbf{K}}_{\varepsilon} = \underline{\mathbf{F}}~ \underline{\mathbf{K}}_{\sigma}~\underline{\mathbf{F}}^{-1} \\
& \underline{\mathbf{C}}' = \underline{\mathbf{K}}_{\sigma}~\underline{\mathbf{C}}~ \underline{\mathbf{K}}_{\varepsilon}^{-1}
\end{split}
\end{equation}
where $\underline{\mathbf{F}}$ is a diagonal matrix with diagonal element values (1, 1, 1, 2, 2, 2). The stress rotation matrix ($\underline{\mathbf{K}}_{\sigma}$) is written in \ref{A1}.

\section{Simulation results and discussions}\label{sec4}
\begin{table}[!t]
\centering
\caption{Material parameters and simulation parameters of Mn-22Cu for improper and proper MT PFM}
\begin{tabular}{ccc}
\toprule[1.5pt]  
parameter & name&  value \\
\midrule[0.75pt]  
$c_{11}$&  elastic constant& 76.588~GPa\\
$c_{12}$&  elastic constant&14.588~GPa\\
$c_{44}$&  elastic constant&31~GPa \\ 
$T_0$& chemical equilibrium temperature& 245~K\\
$Q$& latent heat for improper MT & $4.84\times 10^7$~J/m$^3$  \\
$L^{\eta}$& kinetic coefficient for improper MT& 50~m$^3$/s/J \\
$L^{e}$& kinetic coefficient for proper MT& 1~m$^3$/s/J \\
$\beta^{\eta}$& gradient energy coefficient for improper MT& 2.5$\times 10^{-9}$~J/m \\
$\beta^{e}$& gradient energy coefficient for proper MT& 1$\times 10^{-9}$~J/m \\
$K$& thermal conductivity& 40~J/m/s/K\\
$\rho$& density& 7500~kg/m$^3$\\
$c_\text{v}$& specific heat per unit volume& 2.64$\times$10$^6$~J/m$^3$/K\\
$\alpha$& thermal expansion coefficient& 10$^{-5}$ K$^{-1}$\\
\bottomrule[1.5pt] \label{t1}
\end{tabular}
\end{table}

In this work, we utilize a 2D domain to calculate the eCE for reducing computation  cost. The Mn-22Cu material parameters used for PF simulations are taken from literature~\cite{cui2017three} and summarized in Table~\ref{t1}. 
The material parameters of austenite and martensite in Mn-22Cu are regarded to be the same owing to their negligible differences.
The finite element mesh size should be smaller than the minimum value of interface thickness ($\delta=\sqrt{\beta/2\Delta G^*} \approx 14.8$~nm) between austenite and martensite or between different austenite variants, thus the mesh size is chosen as $\Delta l=10$~nm. 
In the calculation of eCE by indirect and direct method, the uniaxial compressive stress is applied in the $y$-direction (favor the martensitic variant 2 (V2) formation), the left and right boundaries are set mechanically free (inserts in Fig.~\ref{f4}a and \ref{f5}a), and adiabatic boundary conditions are specified by assuming zero heat flux.

\subsection{Comparison of two models  for improper and proper MT}
\begin{figure*}[!t]
\centering
  \includegraphics[width=16cm]{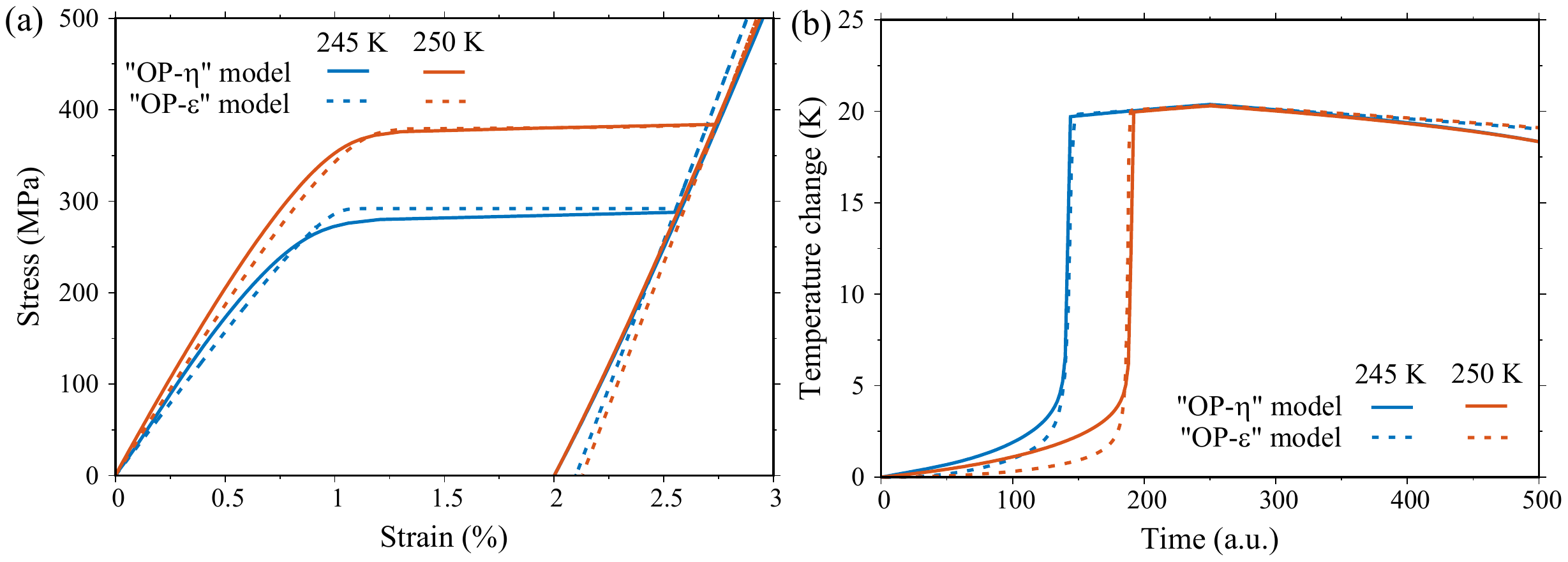}
  \caption{(a) Stress-strain behaviors and (b) temperature change evolution under different temperatures by using the two models. "OP-$\eta$" represents PFM with OPs $\eta$ (improper MT) and "OP-$\varepsilon$" PFM with OPs $\varepsilon_{ii}^0$ (proper MT).} \label{f-eps-eta}
\end{figure*}

The MT behavior and eCE in Mn-22Cu are modelled through two models, the "OP-$\eta$" model using OPs $\eta_I$ for improper MT and "OP-$\varepsilon$" model using OPs $\varepsilon_{ii}^0$ for proper MT.
As shown in Fig.~\ref{f-eps-eta},
under a 500 MPa uniaxial compressive stress, the stress-strain behavior and the temperature change during MT between these two models show good agreement.
The consistency obviously prove the validity and generality of PF simulation in MT and eCE between the two models. The distinction in the temperature change during the beginning of loading is caused by the difference in PF kinetic coefficient $L$.  
In addition, the results of stress-strain behavior and microstructure evolution are consistent between these two models. In fact, Sun et al.~\cite{sun2018phase,sun2019non} have utilized the above two PFMs to simulate the cubic-tetragonal MT in SMA.
Besides, the OPs $\eta_I$ are utilized in PFM built by Khachaturyan et al.~\cite{Khachaturyan1997Three,Artemev2000Phase,artemev2001three} to model the proper and improper MT. 
This shows the correctness and validity of using OPs $\eta$ to model the proper or improper MT at least in term of phenomenological results.

The main discrepancy between the two models is the choice of OPs. 
The eigen strain $\varepsilon_{ij}^0$ for the two models is $\varepsilon_{ij}^{00}(I) \eta_I$ and $\varepsilon_{ii}^0$, respectively. Therefore, the transformation strains in the two models are all 0.02 ($\eta_I=1$ for the former, and $\varepsilon_{ii}^0=-0.02$ for the latter), when the austenite to martensite transformation completely occurs.
For the cubic-tetragonal MT in Mn-22Cu, the two models are equivalent in PF simulation results. Note that the following results of MT behavior and eCE in PF simulation are based on the "OP-$\eta$" model.

\subsection{Benchmark simulation of ferroelastic behavior}
\begin{figure*}[!t]
\centering
  \includegraphics[width=16cm]{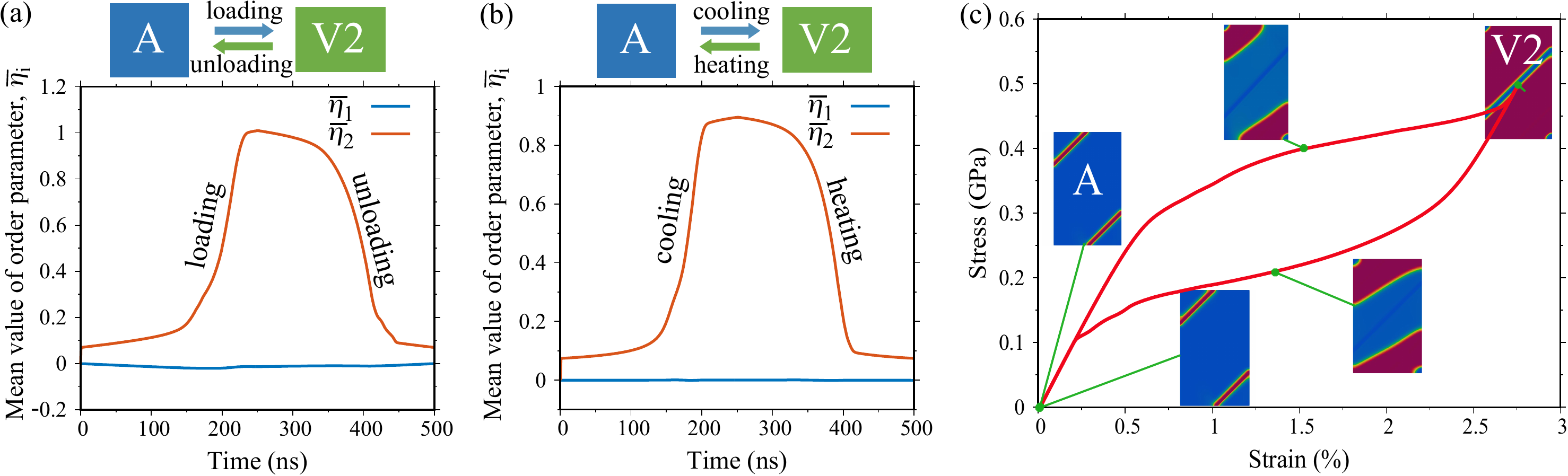}
  \caption{Temporal evolution of order parameters under (a) stress-induced MT and (b) temperature-induced MT. (c) The stress-strain curve of single crystal upon stress loading and unloading at $T=275$~K (above $T_0$). Blue and red are austenite (A) and V2, respectively.} \label{f1}
\end{figure*}

\begin{figure*}[!t]
\centering
  \includegraphics[width=16cm]{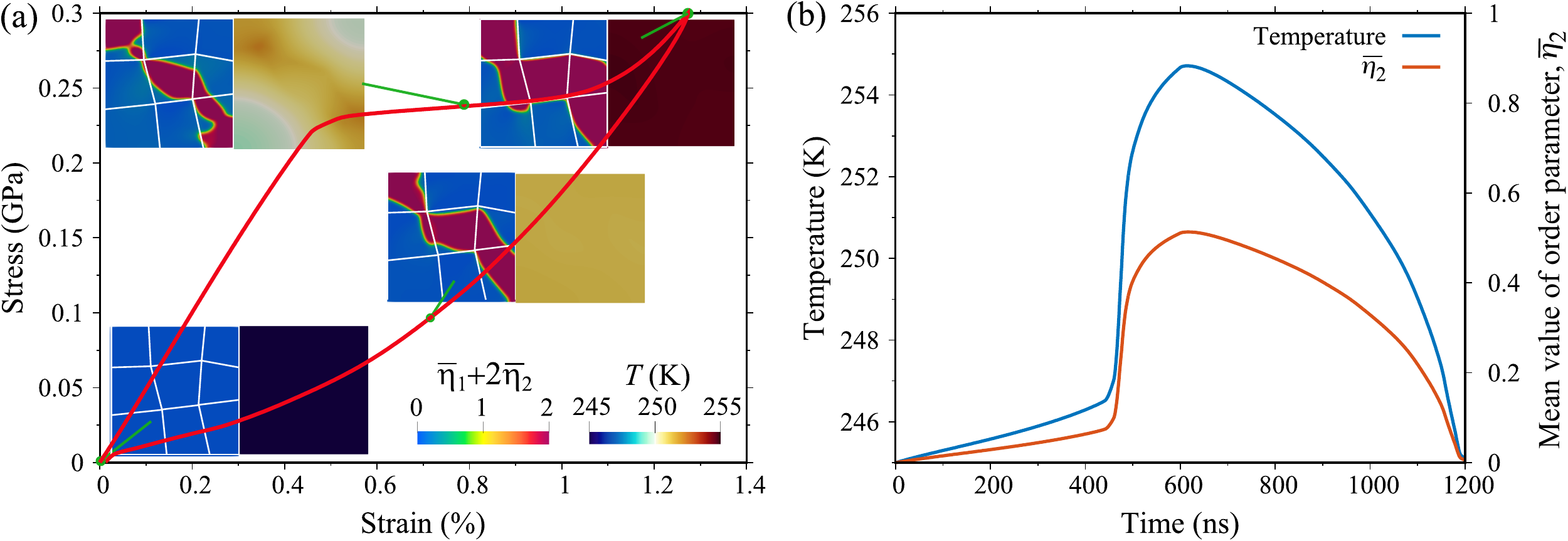}
  \caption{(a) The stress-strain response of polycrystal under $T=245$~K and $\sigma=300$~MPa. Blue and red colors represent A and V2, respectively. (b) Temporal evolution of temperature and $\overline{\eta_I}$.} \label{f2}
\end{figure*}

\begin{figure}[!t]
\centering
  \includegraphics[width=9cm]{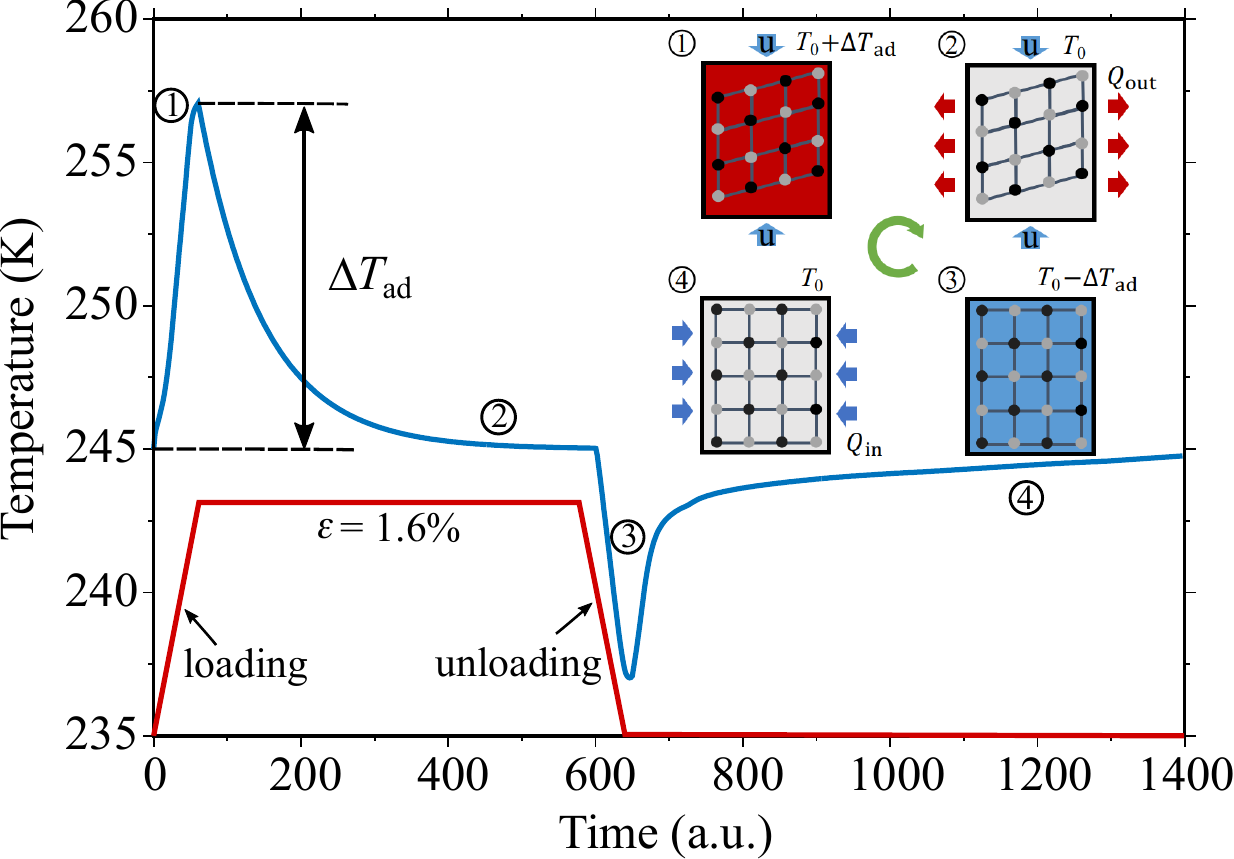}
  \caption{Temperature \textit{vs} time profile and schematics during the thermodynamic cycle.} \label{f3}
\end{figure}

Phase transformation, mainly including stress- and temperature-induced MT, would occur in Mn-22Cu SMAs under external fields, and leads to extraordinary macroscopic behaviors, e.g., the shape memory effect and superelasticity. In order to validate the model, benchmark simulations including that stress- and temperature-induced MT in isothermal (see Fig.~\ref{f1}) and non-isothermal PFM (see Fig.~\ref{f2}), and thermodynamic cycle (see Fig.~\ref{f3}) are carried out.   
For the single grain case in Fig.~\ref{f1}, austenite (A) with a small portion of V2 martensite embryo is set as the initial condition (inset of Fig.~\ref{f1}c).
Figs.~\ref{f1}a and b show the stress- and temperature-induced MT in isothermal PF simulation, respectively.
In Fig.~\ref{f1}a, a large enough compressive stress, along the easy axis of V2, would transform the initial A into V2. The mean value of $\eta_2$ ($\overline{\eta}_2$) increases as the stress is imposed and decreases as the stress is released, indicating the conventional MT during loading and the inverse MT during unloading. 
In Fig.~\ref{f1}b, upon cooling, high-temperature austenite would turn into low-temperature martensite, or \textit{vice versa}.
The superelasticity effect during stress-induced MT is also clearly presented in Fig.~\ref{f1}c.
Upon stress loading, the small martensitic embryo (red stripes) grows to be a large martensitic domain owing to the stress-induced MT. Upon unloading, the austenitic embryo (blue stripes) grows and the original state restores.



Simulating, the superelasticity effect also can be used as a benchmark for validating the non-isothermal PFM for polycrystalline, as shown in Fig.~\ref{f2}. The polycrystalline model size is $1000~\text{nm} \times 1000~\text{nm}$ and has nine grains with random orientations. The maximum compressive stress (along the easy axis of V2) is 300~MPa and the initial temperature is 245~K.
During the loading, the strain increases linearly with the stress at first, corresponding to the elastic response of the austenite. In this regime, there is no MT and temperature and $\overline{\eta}_2$ change a little. As the stress increases to 230~MPa, the stress-induced MT occurs, V2 bands start to grow, and temperatures apparently changes (inset of Fig.~\ref{f2}a).
As shown in Fig.~\ref{f2}b, the temperature and $\overline{\eta}_2$ rapidly rise and then recover to the initial values upon unloading.
The temperature is increased from 245 to 255~K, resulting in $\Delta T_{\text{ad}}=10$~K. The temperature distribution is almost homogeneous because of the high thermal conductivity and no thermal barrier in the domain wall. The temperature is found to be approximately proportional to order parameters, which is consistent with the results reported in~\cite{cui2017three}.


The refrigeration cycle of eCE is schematically shown in the insert of Fig.~\ref{f3}.
When a SMA in the austenitic (cubic) phase is axially stressed or strained, an exothermic austenitic-martensitic transformation occurs, which under adiabatic conditions makes the material heated up (\ding{172} in Fig.~\ref{f3}). This heated material then releases heat to the surroundings and cools down to the ambient temperature (\ding{173} in Fig.~\ref{f3}). When the stress is removed, the crystal structure transforms back to the austenitic phase (\ding{174} in Fig.~\ref{f3}). Finally, the material cools down and is now able to absorb heat from the surroundings (\ding{175} in Fig.~\ref{f3}). 
The simulated temperature \textit{vs} time profile under loading and unloading is also shown in Fig.~\ref{f3}, showing good agreement with the above process from \ding{172} to \ding{175}. These simulation results of MT and thermodynamic cycle indicate that eCE can be soundly handled by the isothermal and non-isothermal PFM.

\subsection{Indirect method}
The indirect method is usually based on the data of superelastic response, such as the stress-strain curves at various temperatures or strain-temperature curves at various stresses. According to the Eqs.~\ref{eq26} and \ref{eq27}, indirect method is used to calculate the isothermal entropy change $\Delta S_{\text{iso}}$ and adiabatic temperature change $\Delta T_{\text{ad}}$ by means of the isothermal PFM. We investigate the eCE under 50, 100, 150, 200, 250, and 300~MPa compressive stress, which is lower than the maximum stress that Mn-22Cu alloy can endure.

\subsubsection{eCE of single crystal}

\begin{figure*}[!t]
\centering
  \includegraphics[width=16cm]{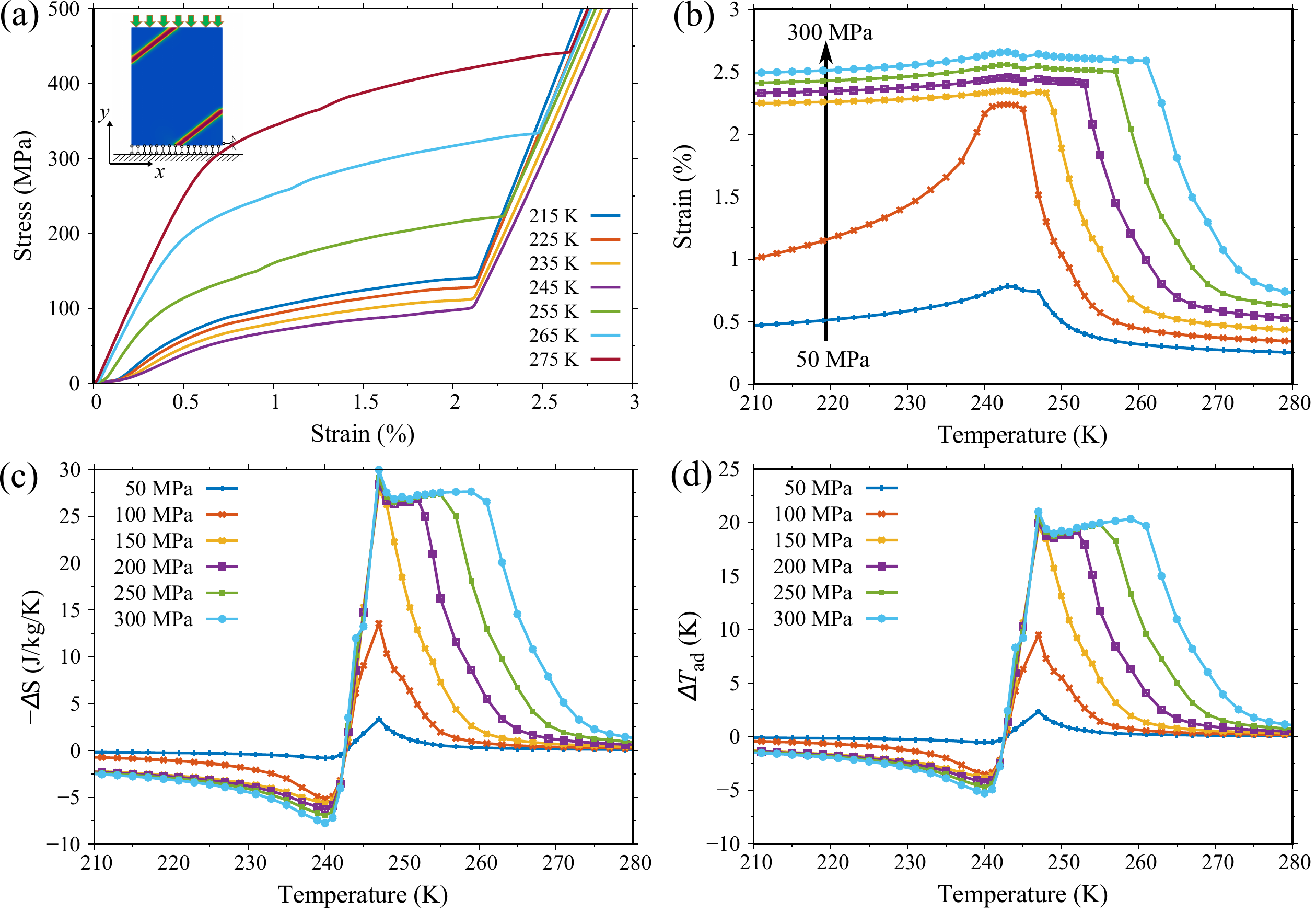}
  \caption{(a) Stress-strain responses of single crystal under 500 MPa compressive stress at various temperatures. (b) Strain-temperature curves under compressive stresses of 50, 100, 150, 200, 250, and 300~MPa. (c) $\Delta S_{\text{iso}}$ and (d) $\Delta T_{\text{ad}}$ calculated by Eq.~\ref{eq26} and \ref{eq27} at various temperatures.} \label{f4}
\end{figure*}

The single crystal model size is $300~\text{nm} \times 500~\text{nm}$.
The typical stress-strain curves at different temperatures from 215 to 275~K are shown in Fig.~\ref{f4}a, which clearly indicates three stages involving elastic stage in austenite, transformation stage, and elastic stage in martensite. A minimum critical transformation stress is found at T = 245~K, because the energy barrier between austenite and martensite vanishes there.
Fig.~\ref{f4}b shows the strain-temperature curves at different compressive stresses from 50 to 300~MPa. During the loading of 100~MPa stress, the strain firstly notably increases and then significantly decreases, indicating that both the inverse and conventional MT occur. By means of the Eqs.~\ref{eq26} and \ref{eq27}, $\Delta S_{\text{iso}}$ and $\Delta T_{\text{ad}}$ can be calculated, as summarized in Figs.~\ref{f4}c and d, respectively. Large $\Delta S_{\text{iso}}$ and $\Delta T_{\text{ad}}$ observably appear near the transformation temperature (245~K). This is caused by the first-order phase transformation which releases or absorbs lots of heat at the transformation temperature.
 
The values of $\Delta S_{\text{max}}$ (J/kg/K) and $\Delta T_{\text{max}}$ (K) at varied stresses are: $\Delta S_{\text{max}}=-3.3$ and $\Delta T_{\text{max}}=2.3$ for $\sigma=50$~MPa, $\Delta S_{\text{max}}=-13.5$ and $\Delta T_{\text{max}}=9.5$ for $\sigma=100$~MPa, $\Delta S_{\text{max}}=-27.8$ and $\Delta T_{\text{max}}=19.5$ for $\sigma=150$~MPa. The maximum $\Delta T_{\text{ad}}$ 9.5~K at 100~MPa is consistent with the experimentally measured value of  11.6~K~\cite{qian2016elastocaloric}. It is found that $\Delta S_{\text{max}}$ and $\Delta T_{\text{max}}$ do not increase further for the stress beyond 150~MPa, because of the already complete MT under 150~MPa stress. However, the operating temperature window is effectively broadened if the stress increases. In addition, there exists an inverse/negative eCE, i.e., positive $\Delta S_{\text{iso}}$ or negative $\Delta T_{\text{ad}}$ under an external stress. This is a consequence of the inverse MT occurring in the case of co-existed austenite and martensite when the ambient temperature is below $T_0$~\cite{bonnot2008elastocaloric,alvarez2017conventional,Xiao2021}. The further explanation in terms of microstructure for the negative eCE will also be discussed in Fig.~\ref{f5}e.

\subsubsection{eCE of polycrystal}

\begin{figure*}[!t]
\centering
  \includegraphics[width=16cm]{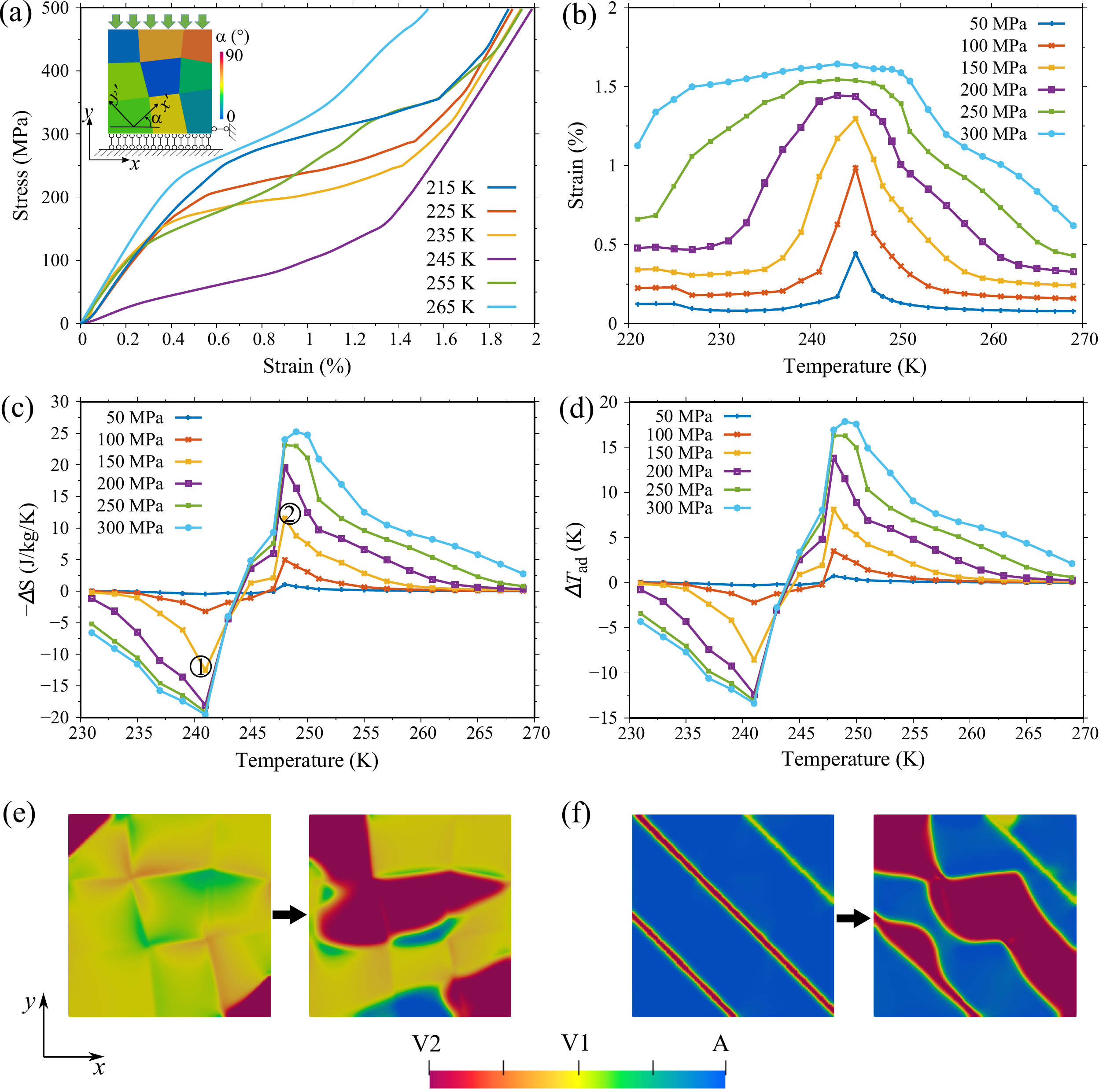}
  \caption{(a) Stress-strain responses of polycrystal under 500~MPa compressive stress at various temperatures. (b) Strain-temperature curves,  (c) $\Delta S_{\text{iso}}$, and (d) $\Delta T_{\text{ad}}$ under various compressive stresses. Phase evolution under 150~MPa at (e) 241~K (\ding{172} in (c)) and (f) 248~K (\ding{173} in (c)). V1, V2, and A are coloured in yellow, red and blue, respectively.} \label{f5}
\end{figure*}

Fig.~\ref{f5} shows the simulation results for eCE of polycrystal.
$\Delta T_{\text{ad}}=3.5$~K for $\sigma=100$~MPa is close to the measured 3.9~K under 4.0$\%$ strain ~\cite{qian2016elastocaloric}.
In general, $\Delta S_{\text{iso}}$ and $\Delta T_{\text{ad}}$ in polycrystalline are lower than that in single crystal. The internal interaction caused by the grain boundary, grains with adverse orientations, and the large negative eCE together contribute to the low $\Delta S_{\text{iso}}$ and $\Delta T_{\text{ad}}$ in polycrystal.
Besides, the large negative $\Delta T_{\text{ad}}$ of polycrystal (-13~K for 200~MPa) in Fig.~\ref{f5}d results from the inverse MT and the local large compressive stress due to the grain boundary, as shown the microstructure evolution in Figs.~\ref{f5}e and f. Below 245 K, the initial phase is martensite, and the inverse MT would occur under a large enough compressive stress. In the case of 150~MPa and 241~K (Fig.~\ref{f5}e), V1 is changed into A at first, and then into V2. This inverse MT absorbs an amount of heat, resulting in the large inverse eCE. At 248~K (Fig.~\ref{f5}f), the initial austenite phase is transformed into V2 accompanied with a small portion of V1.

\subsection{Direct method} 
In experiment or numerical simulation, the straightforward way to assess eCE is the direct method, which directly measures the $\Delta T_{\text{ad}}$ during loading and unloading. Ciss{\'e} et al.~\cite{cisse2020elastocaloric} and Sun et al.~\cite{sun2019non} used the non-isothermal PFM to directly calculate $\Delta T_{\text{ad}}$ in CuAl$_{11}$Be$_2$ and FePd SMAs, respectively. Here we use the modified non-isothermal PFM to evaluate eCE in Mn-22Cu
alloys based the direct method. The applied compressive stress is 100~MPa, the loading rate is 0.2~MPa/ns, and other conditions are consistent with the indirect method for a comparison.

\subsubsection{eCE of single crystal}
\begin{figure*}[!t]
\centering
  \includegraphics[width=16cm]{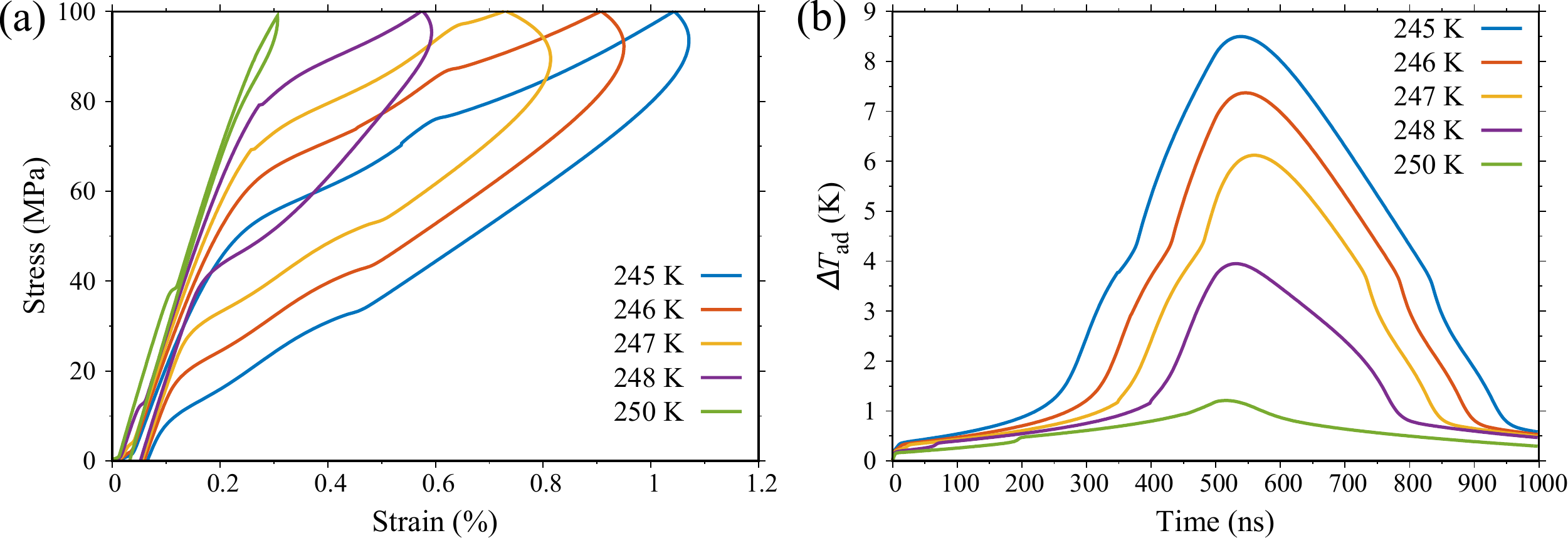}
  \caption{(a) Stress-strain curves and (b) $\Delta T_{\text{ad}}$ of single crystal under 100~MPa compressive stress at different temperatures.}\label{f6}
\end{figure*}

For single crystal, the strain-stress response and the adiabatic temperature change calculated by the direct method are shown in Fig.~\ref{f6}. It can be found in Fig.~\ref{f6}a that the sample almost stays at the linear elastic regime at 250~K greater than the equilibrium temperature. As the initial temperature decreases to 245~K, the critical transformation stress decreases and the area with MT increases, leading to in a larger $\Delta T_{\text{ad}}$ (Fig.~\ref{f6}b). The transformation strain is 0.8$\%$ at 245 K under 100~MPa compressive stress and the corresponding $\Delta T_{\text{ad}}$ is around 8.5~K. $\Delta T_{\text{ad}}$ in Fig.~\ref{f6}b shows a strong temperature dependence. In other words, controlling the ambient temperature to obtain high transformation strain is a feasible neat idea to improve the eCE of SMAs.

\subsubsection{eCE of polycrystal}
\begin{figure*}[!t]
\centering
  \includegraphics[width=16cm]{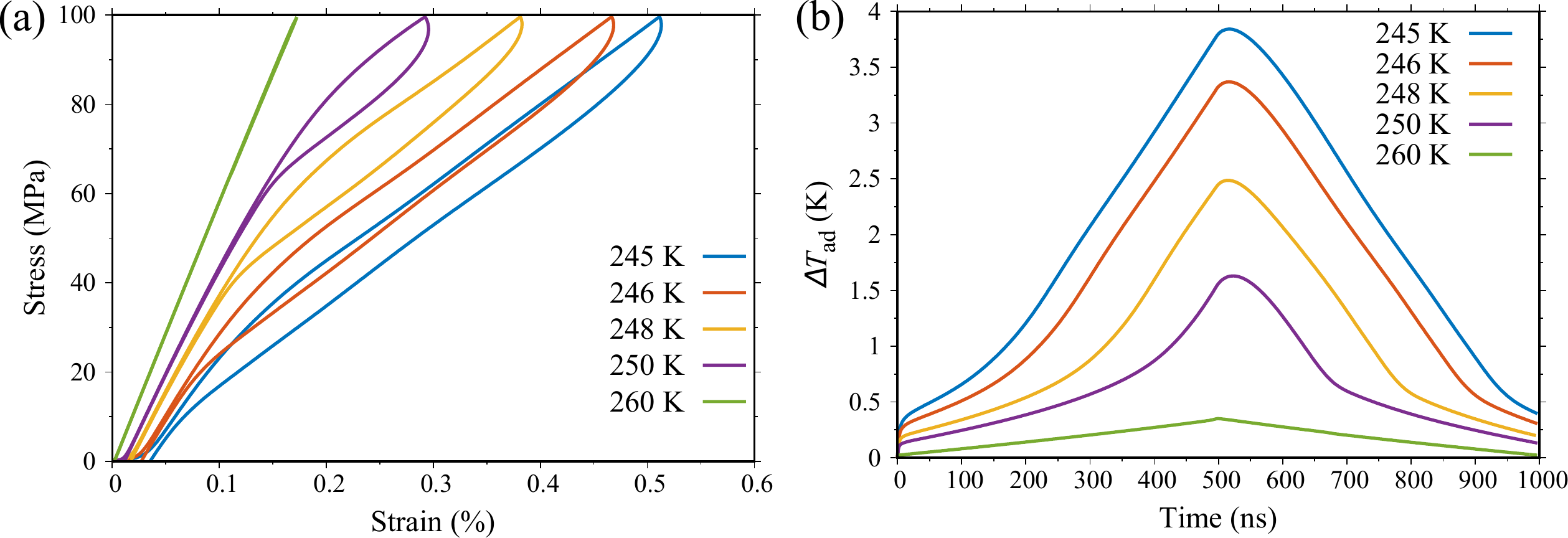}
  \caption{(a) Stress-strain curves and (b) $\Delta T_{\text{ad}}$ of polycrystal under 100~MPa compressive stress at different temperatures.}\label{f7}
\end{figure*}

The settings of polycrystalline model are consistent with those in the indirect method. Fig.~\ref{f7}a and b show the stress-strain response and $\Delta T_{\text{ad}}$ at different temperatures, respectively. Under the same stress, the maximum transformation strain and $\Delta T_{\text{ad}}$ occurs at $T$=~245 K and the corresponding $\Delta T_{\text{ad}}$ is around 3.8~K. Polycrystal is more closer to the microstructure of experimental samples and thus its maximum $\Delta T_{\text{ad}}$ of 3.8~K by the direct method and 3.5~K by the indirect method is more closer to the experimental data (3.9~K)~\cite{qian2016elastocaloric}.
Compared to the maximum $\Delta T_{\text{ad}}$ of 8.5~K in single crystal, the smaller $\Delta T_{\text{ad}}$ in polycrystal is probably a consequence of domain wall and adverse orientation grain in the polycrystalline model. The grain misorientation may hinder the martensitic nucleation and growth, as shown in Fig.~\ref{f2}a.
In addition, comparing Fig.~\ref{f6}a with Fig.~\ref{f7}a, it is obvious that a smaller hysteresis exists in polycrystal.

\subsection{Indirect \textit{vs} direct method}
\begin{figure*}[!t]
\centering
  \includegraphics[width=16cm]{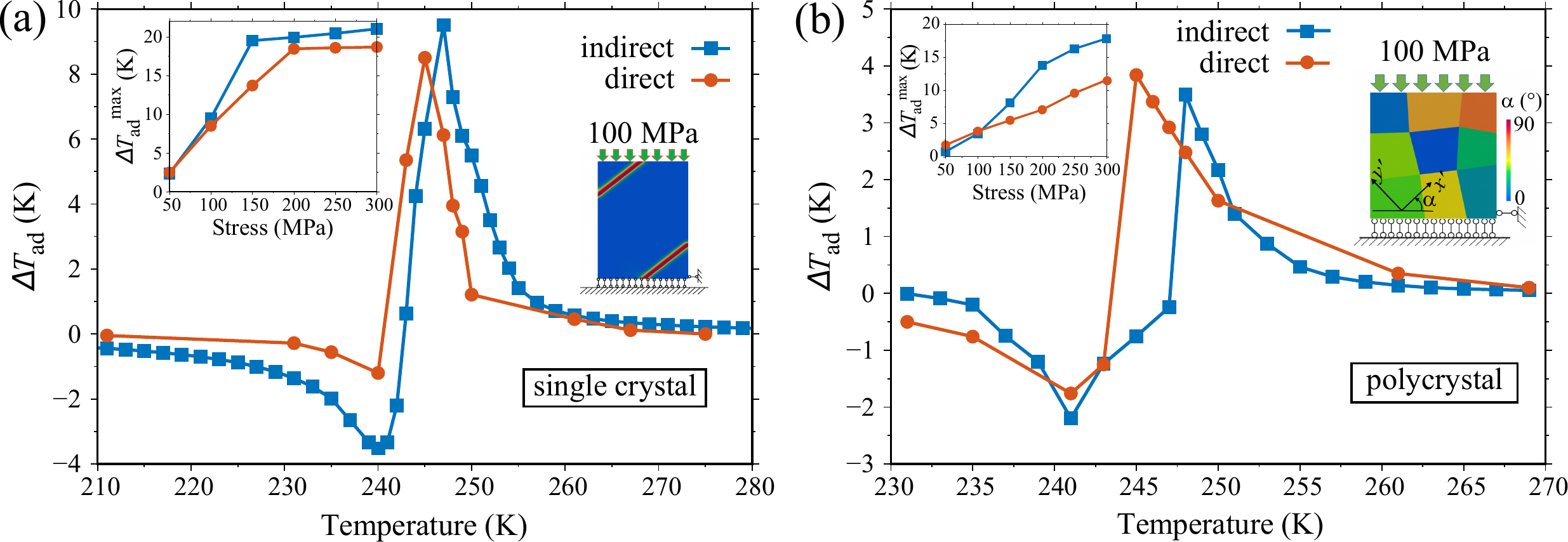}
  \caption{$\Delta T_{\text{ad}}$-$T$ curves of (a) single crystal and (b) polycrystal calculated by the indirect and direct method under a compressive stress of 100~MPa. The insets show the $\Delta T_{\text{ad}}^{\text{max}}$ as a function of compressive stress.} \label{f8}
\end{figure*}

In order to compare the indirect and direct method which are based on PFM, in Fig.~\ref{f8} we present the $\Delta T_{\text{ad}}$-$T$ curves calculated by both methods under 100~MPa compressive stress. It is clear that the overall trend of $\Delta T_{\text{ad}}$ varying with $T$ is consistent for both methods. For single crystal in Fig.~\ref{f8}a, $\Delta T_{\text{ad}}$ calculated by the two methods at different ambient temperatures is close. Under 100~MPa, the maximum $\Delta T_{\text{ad}}$ ($\Delta T_{\text{ad}}^{\text{max}}$) calculated by indirect and direct method is 9.5 and 8.5~K (3.5 and 3.8~K) for single crystal (polycrystal), respectively. This indicates the consistency between indirect and direct method and the reliability of calculating eCE by PF simulations.
However, the peak temperature corresponding to $\Delta T_{\text{ad}}^{\text{max}}$ calculated by the two methods is slightly different. For the direct method, the peak temperature is around 245~K (the chemical equilibrium temperature), which is lower than the indirect method (247~K for single crystal and 248~K for polycrystal).


The insets in Fig.~\ref{f8} present the $\Delta T_{\text{ad}}^{\text{max}}$ as a function of the applied compressive stress. It can be seen that eCE can be significantly tuned by the uniaxial compressive stress. Results from the direct method is essentially in agreement with those from the indirect method when the stress is less than 150~MPa, with a discrepancy of $\Delta T_{\text{ad}}^{\text{max}}$ within 10$\%$. At higher compressive stress (above 150~MPa), $\Delta T_{\text{ad}}^{\text{max}}$ calculated by the indirect method is apparently larger than that form the direct one. This could be attributed to the temperature-induced MT in the non-isothermal PFM simulation, i.e., the relatively large temperature increase at high stress could in turn remarkably hinder the austenite-martensite transition and thus lower the temperature~\cite{chen2021improved}. Cheng et al.~\cite{Cheng2020} also report the similar phenomena in electrocaloric effect. In addition, $\Delta T_{\text{ad}}/\sigma_{\text{max}}$ is usually utilized as a parameter to evaluate the field normalized caloric effect, also known as the specific adiabatic temperature~\cite{xu2016giant}. It is found that $\Delta T_{\text{ad}}/\sigma_{\text{max}}$ is 0.131 and 0.091 K/MPa for single crystal (0.054 and 0.037 K/MPa for polycrystal) by the indirect and direct method, respectively. A large $\Delta T_{\text{ad}}/\sigma_{\text{max}}$ would be beneficial to enhance the overall performance and efficiency of the elastocaloric refrigerator.


The agreement between the calculated and experimental results is satisfactory, implying that the direct method based on the non-isothermal PFM and the direct method based on isothermal PFM are reliable for the calculation and evaluation of eCE in SMAs. Nevertheless, the computational cost and the available information on eCE are also critical factors to be considered. The great advantage of the indirect method is that all the $\Delta S_{\text{iso}}$, $\Delta T_{\text{ad}}$, and refrigerating capacity can be readily obtained. But the indirect method depends on the temperature dependent strain-stress curves with small temperature intervals for an accurate integration, leading to heavy computation. If one is only interested in $\Delta T_{\text{ad}}$, the direct method is the best choice since it doe not require the overall stress-strain-temperature data and thus is computationally efficient.

\section{Conclusion}\label{sec5}
In summary, we have developed a thermodynamically consistent non-isothermal PFM for the simulation of eCE. The PFM couples MT with mechanics and heat transfer to evaluate eCE by using the indirect and direct method. The model considering both improper and proper MT is derived from a thermodynamic framework which invokes the microforce theory accommodating non-local effects, thermodynamic laws, and the Coleman--Noll procedure. In order to avoid the problematic calculation due to the non-differentiable energy barrier function across the transformation temperature and consider the possibly non-sharp transition in real materials, the austenite-martensite transition energy barrier in PFM is introduced as a smooth function of temperature by using the hyperbolic tangent function.
On one hand, all constitutive relations are represented in terms of a thermodynamic potential. Therefore, the PF modelling work is reduced to the design of a proper form of the thermodynamic potential. On the the hand, the framework automatically satisfies the first, second, and third laws of thermodynamics.

After being numerically implemented by the finite element method, the developed PFM is demonstrated to be capable of recapturing the microstructure response and calculating eCE via both indirect and direct method in a model material (i.e., Mn-22Cu SMA) under an external loading. Under a compressive stress of 100~MPa, $\Delta T_{\text{ad}}^{\text{max}}$ calculated by the indirect and direct method is 9.5 and 8.5~K for single crystal (3.5 and 3.8~K for polycrystal), respectively. A large $\Delta T_{\text{ad}}$ exists in single crystal, but the working temperature window is narrow and can be improved by increasing the compressive stress. Besides, negative eCE caused by the inverse MT is found, especially for polycrystal, which reduces $\Delta T_{\text{ad}}^{\text{max}}$.
Overall, the direct method based on the non-isothermal PFM and the direct method based on isothermal PFM are reliable for the calculation and evaluation of eCE in SMAs. But there are still some differences between these two methods.
$\Delta T_{\text{ad}}$ calculated by the indirect and direct method shows tiny discrepancy (within 10$\%$) under a low stress ($\leq$150~MPa). 
At higher stress, $\Delta T_{\text{ad}}^{\text{max}}$ from the indirect method is apparently larger than that from the direct one, mainly owing to the fact that the relatively large temperature increase at high stress in the non-isothermal PFM could in turn remarkably hinder the austenite-martensite transition and thus lower the temperature.
The indirect method is computational expensive, but can yield all the $\Delta S_{\text{iso}}$, $\Delta T_{\text{ad}}$, and refrigerating capacity. The direct method is computationally efficient, but only yields $\Delta T_{\text{ad}}$. The developed non-isothermal PFM along with the indirect and direct method could be a computational toolkit to unveil strategies for designing high-performance elastocaloric devices.

\section*{CRediT authorship contribution statement}
\textbf{Wei Tang}: Formal analysis, Investigation, Visualization, Data curation, Writing -- original draft. \textbf{Qihua Gong}: Conceptualization, Resources, Supervision, Investigation, Data curation, Writing -- original draft \& review \& editing. \textbf{Min Yi}: Conceptualization, Resources, Supervision, Project administration, Funding acquisition, Writing -- original draft \& review \& editing. \textbf{Bai-Xiang Xu}: Conceptualization, Supervision, Writing -- original draft \& review.  \textbf{Long-Qing Chen}: Conceptualization, Supervision, Writing -- original draft \& review.

\section*{Declaration of competing interest}
The authors declare that they have no known competing financial interests or personal relationships that could have appeared to influence the work reported in this paper.

\section*{Acknowledgements}
The authors acknowledge the support from National Natural Science Foundation of China (NSFC 12272173, 11902150), 15$^\text{th}$ Thousand Youth Talents Program of China, Fundamental Research Funds for the Central Universities (1001-XAC21021), the Research Fund of State Key Laboratory of Mechanics and Control of Mechanical Structures (MCMS-I-0419G01), and a project Funded by the Priority Academic Program Development of Jiangsu Higher Education Institutions.

\bibliographystyle{elsarticle-num}
\bibliography{riaibib}

\begin{thebibliography}{10}
\expandafter\ifx\csname url\endcsname\relax
  \def\url#1{\texttt{#1}}\fi
\expandafter\ifx\csname urlprefix\endcsname\relax\def\urlprefix{URL }\fi
\expandafter\ifx\csname href\endcsname\relax
  \def\href#1#2{#2} \def\path#1{#1}\fi

\bibitem{wada2001giant}
H.~Wada, Y.~Tanabe, Giant magnetocaloric effect of {MnAs$_{1-x}$Sb$_x$},
  Applied Physics Letters 79~(20) (2001) 3302--3304.
\newblock \href {https://doi.org/10.1063/1.1419048}
  {\path{doi:10.1063/1.1419048}}.

\bibitem{phan2007review}
M.-H. Phan, S.-C. Yu, Review of the magnetocaloric effect in manganite
  materials, Journal of Magnetism and Magnetic Materials 308~(2) (2007)
  325--340.
\newblock \href {https://doi.org/10.1016/j.jmmm.2006.07.025}
  {\path{doi:10.1016/j.jmmm.2006.07.025}}.

\bibitem{gschneidnerjr2005recent}
K.~A. GschneidnerJr, V.~Pecharsky, A.~Tsokol, Recent developments in
  magnetocaloric materials, Reports on Progress in Physics 68~(6) (2005)
  14--79.
\newblock \href {https://doi.org/10.1088/0034-4885/68/6/R04}
  {\path{doi:10.1088/0034-4885/68/6/R04}}.

\bibitem{mischenko2006giant}
A.~Mischenko, Q.~Zhang, J.~Scott, R.~Whatmore, N.~Mathur, Giant electrocaloric
  effect in thin-film {PbZr$_{0.95}$Ti$_{0. 05}$O$_3$}, Science 311~(57) (2006)
  1270--1271.
\newblock \href {https://doi.org/10.1126/science.1123811}
  {\path{doi:10.1126/science.1123811}}.

\bibitem{scott2011electrocaloric}
J.~Scott, Electrocaloric materials, Annual Review of Materials Research 41
  (2011) 229--240.
\newblock \href {https://doi.org/10.1146/annurev-matsci-062910-100341}
  {\path{doi:10.1146/annurev-matsci-062910-100341}}.

\bibitem{manosa2010giant}
L.~Ma{\~n}osa, D.~Gonz{\'a}lez-Alonso, A.~Planes, E.~Bonnot, M.~Barrio, J.-L.
  Tamarit, S.~Aksoy, M.~Acet, Giant solid-state barocaloric effect in the
  {Ni--Mn--In} magnetic shape-memory alloy, Nature materials 9~(6) (2010)
  478--481.
\newblock \href {https://doi.org/10.1038/NMAT2731}
  {\path{doi:10.1038/NMAT2731}}.

\bibitem{lloveras2015giant}
P.~Lloveras, E.~Stern-Taulats, M.~Barrio, J.-L. Tamarit, S.~Crossley, W.~Li,
  V.~Pomjakushin, A.~Planes, L.~Ma{\~n}osa, N.~Mathur, et~al., Giant
  barocaloric effects at low pressure in ferrielectric ammonium sulphate,
  Nature communications 6~(1) (2015) 1--6.
\newblock \href {https://doi.org/10.1038/ncomms9801}
  {\path{doi:10.1038/ncomms9801}}.

\bibitem{bonnot2008elastocaloric}
E.~Bonnot, R.~Romero, L.~Ma{\~n}osa, E.~Vives, A.~Planes, Elastocaloric effect
  associated with the martensitic transition in shape-memory alloys, Physical
  Review Letters 100~(12) (2008) 125901.
\newblock \href {https://doi.org/10.1103/PhysRevLett.100.125901}
  {\path{doi:10.1103/PhysRevLett.100.125901}}.

\bibitem{tuvsek2015elastocaloricni}
J.~Tu{\v{s}}ek, K.~Engelbrecht, L.~P. Mikkelsen, N.~Pryds, Elastocaloric effect
  of {Ni-Ti} wire for application in a cooling device, Journal of Applied
  Physics 117~(12) (2015) 124901.
\newblock \href {https://doi.org/10.1063/1.4913878}
  {\path{doi:10.1063/1.4913878}}.

\bibitem{manosa2013large}
L.~Ma{\~n}osa, S.~Jarque-Farnos, E.~Vives, A.~Planes, Large temperature span
  and giant refrigerant capacity in elastocaloric {Cu-Zn-Al} shape memory
  alloys, Applied Physics Letters 103~(21) (2013) 211904.
\newblock \href {https://doi.org/10.1063/1.4832339}
  {\path{doi:10.1063/1.4832339}}.

\bibitem{tuvsek2015elastocaloric}
J.~Tu{\v{s}}ek, K.~Engelbrecht, R.~Mill{\'a}n-Solsona, L.~Ma{\~n}osa, E.~Vives,
  L.~P. Mikkelsen, N.~Pryds, The elastocaloric effect: a way to cool
  efficiently, Advanced Energy Materials 5~(13) (2015) 1500361.
\newblock \href {https://doi.org/10.1002/aenm.201500361}
  {\path{doi:10.1002/aenm.201500361}}.

\bibitem{cui2012demonstration}
J.~Cui, Y.~Wu, J.~Muehlbauer, Y.~Hwang, R.~Radermacher, S.~Fackler, M.~Wuttig,
  I.~Takeuchi, {Demonstration of high efficiency elastocaloric cooling with
  large $\Delta$T using NiTi wires}, Applied Physics Letters 101~(7) (2012)
  073904.
\newblock \href {https://doi.org/10.1063/1.4746257}
  {\path{doi:10.1063/1.4746257}}.

\bibitem{ovzbolt2014electrocaloric}
M.~O{\v{z}}bolt, A.~Kitanovski, J.~Tu{\v{s}}ek, A.~Poredo{\v{s}},
  Electrocaloric vs. magnetocaloric energy conversion, International Journal of
  Refrigeration 37 (2014) 16--27.
\newblock \href {https://doi.org/10.1016/j.ijrefrig.2013.07.001}
  {\path{doi:10.1016/j.ijrefrig.2013.07.001}}.

\bibitem{qian2016review}
S.~Qian, Y.~Geng, Y.~Wang, J.~Ling, Y.~Hwang, R.~Radermacher, I.~Takeuchi,
  J.~Cui, A review of elastocaloric cooling: Materials, cycles and system
  integrations, International Journal of Refrigeration 64 (2016) 1--19.
\newblock \href {https://doi.org/10.1016/j.ijrefrig.2015.12.001}
  {\path{doi:10.1016/j.ijrefrig.2015.12.001}}.

\bibitem{chen2002phase}
L.-Q. Chen, Phase-field models for microstructure evolution, Annual review of
  materials research 32~(1) (2002) 113--140.
\newblock \href {https://doi.org/10.1146/annurev.matsci.32.112001.132041}
  {\path{doi:10.1146/annurev.matsci.32.112001.132041}}.

\bibitem{chen2022classical}
L.-Q. Chen, Y.~Zhao, From classical thermodynamics to phase-field method,
  Progress in Materials Science 124 (2022) 100868.
\newblock \href {https://doi.org/10.1016/j.pmatsci.2021.100868}
  {\path{doi:10.1016/j.pmatsci.2021.100868}}.

\bibitem{tanaka1986thermomechanical}
K.~Tanaka, A thermomechanical sketch of shape memory effect: one-dimensional
  tensile behavior, Res. Mechanica 18~(3) (1986) 251--263.

\bibitem{ossmer2014evolution}
H.~Ossmer, F.~Lambrecht, M.~G{\"u}ltig, C.~Chluba, E.~Quandt, M.~Kohl,
  Evolution of temperature profiles in {TiNi} films for elastocaloric cooling,
  Acta Materialia 81 (2014) 9--20.
\newblock \href {https://doi.org/10.1016/j.actamat.2014.08.006}
  {\path{doi:10.1016/j.actamat.2014.08.006}}.

\bibitem{krevet2013evolution}
B.~Krevet, V.~Pinneker, M.~Rhode, C.~Bechthold, E.~Quandt, M.~Kohl, {Evolution
  of temperature profiles during stress-induced transformation in NiTi thin
  films}, in: Materials Science Forum, Vol. 738, Trans Tech Publ, 2013, pp.
  287--291.
\newblock \href {https://doi.org/10.4028/www.scientific.net/MSF.738-739.287}
  {\path{doi:10.4028/www.scientific.net/MSF.738-739.287}}.

\bibitem{tuvsek2016understanding}
J.~Tu{\v{s}}ek, K.~Engelbrecht, L.~Ma{\~n}osa, E.~Vives, N.~Pryds,
  Understanding the thermodynamic properties of the elastocaloric effect
  through experimentation and modelling, Shape Memory and Superelasticity 2~(4)
  (2016) 317--329.
\newblock \href {https://doi.org/10.1007/s40830-016-0094-8}
  {\path{doi:10.1007/s40830-016-0094-8}}.

\bibitem{luo2017modeling}
D.~Luo, Y.~Feng, P.~Verma, Modeling and analysis of an integrated solid state
  elastocaloric heat pumping system, Energy 130 (2017) 500--514.
\newblock \href {https://doi.org/10.1016/j.energy.2017.05.008}
  {\path{doi:10.1016/j.energy.2017.05.008}}.

\bibitem{qian2015thermodynamics}
S.~Qian, J.~Ling, Y.~Hwang, R.~Radermacher, I.~Takeuchi, Thermodynamics cycle
  analysis and numerical modeling of thermoelastic cooling systems,
  International Journal of Refrigeration 56 (2015) 65--80.
\newblock \href {https://doi.org/10.1016/j.ijrefrig.2015.04.001}
  {\path{doi:10.1016/j.ijrefrig.2015.04.001}}.

\bibitem{qian2017mechanism}
S.~Qian, L.~Yuan, J.~Yu, G.~Yan, {The mechanism of $\Delta$T variation in
  coupled heat transfer and phase transformation for elastocaloric materials
  and its application in materials characterization}, Applied Physics Letters
  111~(22) (2017) 223902.
\newblock \href {https://doi.org/10.1063/1.5001971}
  {\path{doi:10.1063/1.5001971}}.

\bibitem{qian2017numerical}
S.~Qian, L.~Yuan, J.~Yu, G.~Yan, Numerical modeling of an active elastocaloric
  regenerator refrigerator with phase transformation kinetics and the matching
  principle for materials selection, Energy 141 (2017) 744--756.
\newblock \href {https://doi.org/10.1016/j.energy.2017.09.116}
  {\path{doi:10.1016/j.energy.2017.09.116}}.

\bibitem{yu2020modeling}
C.~Yu, T.~Chen, H.~Yin, G.~Kang, D.~Fang, {Modeling the anisotropic
  elastocaloric effect of textured NiMnGa ferromagnetic shape memory alloys},
  International Journal of Solids and Structures 191 (2020) 509--528.
\newblock \href {https://doi.org/10.1016/j.ijsolstr.2019.12.020}
  {\path{doi:10.1016/j.ijsolstr.2019.12.020}}.

\bibitem{zhou2020modeling}
T.~Zhou, G.~Kang, H.~Yin, C.~Yu, Modeling the two-way shape memory and
  elastocaloric effects of bamboo-grained oligocrystalline shape memory alloy
  microwire, Acta Materialia 198 (2020) 10--24.
\newblock \href {https://doi.org/10.1016/j.actamat.2020.07.057}
  {\path{doi:10.1016/j.actamat.2020.07.057}}.

\bibitem{chen2004phase}
L.-Q. Chen, S.~Hu, Phase-field method applied to strain-dominated
  microstructure evolution during solid-state phase transformations, Continuum
  Scale Simulation of Engineering Materials:
  Fundamentals--Microstructures--Process Applications (2004) 271--296\href
  {https://doi.org/10.1002/3527603786.ch11}
  {\path{doi:10.1002/3527603786.ch11}}.

\bibitem{wang1997three}
Y.~Wang, A.~Khachaturyan, Three-dimensional field model and computer modeling
  of martensitic transformations, Acta Materialia 45~(2) (1997) 759--773.
\newblock \href {https://doi.org/10.1016/S1359-6454(96)00180-2}
  {\path{doi:10.1016/S1359-6454(96)00180-2}}.

\bibitem{wen2000phase}
Y.~Wen, Y.~Wang, L.-Q. Chen, Phase-field simulation of domain structure
  evolution during a coherent hexagonal-to-orthorhombic transformation,
  Philosophical Magazine A 80~(9) (2000) 1967--1982.
\newblock \href {https://doi.org/10.1080/01418610008212146}
  {\path{doi:10.1080/01418610008212146}}.

\bibitem{seol2003cubic}
D.~Seol, S.~Hu, Y.~Li, L.~Chen, K.~Oh, Cubic to tetragonal martensitic
  transformation in a thin film elastically constrained by a substrate, Metals
  and Materials International 9~(3) (2003) 221--226.
\newblock \href {https://doi.org/10.1007/BF03027039}
  {\path{doi:10.1007/BF03027039}}.

\bibitem{mamivand2013phase}
M.~Mamivand, M.~A. Zaeem, H.~El~Kadiri, L.-Q. Chen, Phase field modeling of the
  tetragonal-to-monoclinic phase transformation in zirconia, Acta Materialia
  61~(14) (2013) 5223--5235.
\newblock \href {https://doi.org/10.1016/j.actamat.2013.05.015}
  {\path{doi:10.1016/j.actamat.2013.05.015}}.

\bibitem{levitas2002three}
V.~I. Levitas, D.~L. Preston, {Three-dimensional Landau theory for multivariant
  stress-induced martensitic phase transformations. I.
  Austenite$\ensuremath{\leftrightarrow}$martensite}, Physical Review B 66~(13)
  (2002) 134206.
\newblock \href {https://doi.org/10.1103/PhysRevB.66.134206}
  {\path{doi:10.1103/PhysRevB.66.134206}}.

\bibitem{levitas2002three2}
V.~I. Levitas, D.~L. Preston, {Three-dimensional Landau theory for multivariant
  stress-induced martensitic phase transformations. II. Multivariant phase
  transformations and stress space analysis}, Physical Review B 66~(13) (2002)
  134207.
\newblock \href {https://doi.org/levitas2002three}
  {\path{doi:levitas2002three}}.

\bibitem{levitas2003three}
V.~I. Levitas, D.~L. Preston, D.-W. Lee, {Three-dimensional Landau theory for
  multivariant stress-induced martensitic phase transformations. III.
  Alternative potentials, critical nuclei, kink solutions, and dislocation
  theory}, Physical Review B 68~(13) (2003) 134201.
\newblock \href {https://doi.org/10.1103/PhysRevB.68.134201}
  {\path{doi:10.1103/PhysRevB.68.134201}}.

\bibitem{levitas2010surface}
V.~I. Levitas, M.~Javanbakht, Surface tension and energy in multivariant
  martensitic transformations: phase-field theory, simulations, and model of
  coherent interface, Physical Review Letters 105~(16) (2010) 165701.
\newblock \href {https://doi.org/10.1103/PhysRevLett.105.165701}
  {\path{doi:10.1103/PhysRevLett.105.165701}}.

\bibitem{levitas2013thermodynamically}
V.~I. Levitas, Thermodynamically consistent phase field approach to phase
  transformations with interface stresses, Acta Materialia 61~(12) (2013)
  4305--4319.
\newblock \href {https://doi.org/10.1016/j.actamat.2013.03.034}
  {\path{doi:10.1016/j.actamat.2013.03.034}}.

\bibitem{levitas2009displacive}
V.~I. Levitas, V.~A. Levin, K.~M. Zingerman, E.~I. Freiman, Displacive phase
  transitions at large strains: phase-field theory and simulations, Physical
  Review Letters 103~(2) (2009) 025702.
\newblock \href {https://doi.org/10.1103/PhysRevLett.103.025702}
  {\path{doi:10.1103/PhysRevLett.103.025702}}.

\bibitem{levitas2013phase}
V.~I. Levitas, Phase-field theory for martensitic phase transformations at
  large strains, International Journal of Plasticity 49 (2013) 85--118.
\newblock \href {https://doi.org/10.1016/j.ijplas.2013.03.002}
  {\path{doi:10.1016/j.ijplas.2013.03.002}}.

\bibitem{hou2019fatigue}
H.~Hou, E.~Simsek, T.~Ma, N.~S. Johnson, S.~Qian, C.~Ciss{\'e}, D.~Stasak,
  N.~Al~Hasan, L.~Zhou, Y.~Hwang, et~al., Fatigue-resistant high-performance
  elastocaloric materials made by additive manufacturing, Science 366~(6469)
  (2019) 1116--1121.
\newblock \href {https://doi.org/10.1126/science.aax7616}
  {\path{doi:10.1126/science.aax7616}}.

\bibitem{cisse2020elastocaloric}
C.~Ciss{\'e}, M.~A. Zaeem, On the elastocaloric effect in cualbe shape memory
  alloys: A quantitative phase-field modeling approach, Computational Materials
  Science 183 (2020) 109808.
\newblock \href {https://doi.org/10.1016/j.commatsci.2020.109808}
  {\path{doi:10.1016/j.commatsci.2020.109808}}.

\bibitem{cisse2020asymmetric}
C.~Ciss{\'e}, M.~A. Zaeem, An asymmetric elasto-plastic phase-field model for
  shape memory effect, pseudoelasticity and thermomechanical training in
  polycrystalline shape memory alloys, Acta Materialia 201 (2020) 580--595.
\newblock \href {https://doi.org/10.1016/j.actamat.2020.10.034}
  {\path{doi:10.1016/j.actamat.2020.10.034}}.

\bibitem{CISSE2021109898}
C.~Cissé, M.~{Asle Zaeem}, {Design of NiTi-based shape memory microcomposites
  with enhanced elastocaloric performance by a fully thermomechanical coupled
  phase-field model}, Materials \& Design 207 (2021) 109898.
\newblock \href {https://doi.org/https://doi.org/10.1016/j.matdes.2021.109898}
  {\path{doi:https://doi.org/10.1016/j.matdes.2021.109898}}.

\bibitem{levitas2015multiphase}
V.~I. Levitas, A.~M. Roy, Multiphase phase field theory for temperature-and
  stress-induced phase transformations, Physical Review B 91~(17) (2015)
  174109.
\newblock \href {https://doi.org/10.1103/PhysRevB.91.174109}
  {\path{doi:10.1103/PhysRevB.91.174109}}.

\bibitem{cui2017three}
S.~Cui, J.~Wan, X.~Zuo, N.~Chen, J.~Zhang, Y.~Rong, Three-dimensional,
  non-isothermal phase-field modeling of thermally and stress-induced
  martensitic transformations in shape memory alloys, International Journal of
  Solids and Structures 109 (2017) 1--11.
\newblock \href {https://doi.org/10.1016/j.ijsolstr.2017.01.001}
  {\path{doi:10.1016/j.ijsolstr.2017.01.001}}.

\bibitem{sun2018phase}
Y.~Sun, J.~Luo, J.~Zhu, Phase field study of the microstructure evolution and
  thermomechanical properties of polycrystalline shape memory alloys: Grain
  size effect and rate effect, Computational Materials Science 145 (2018)
  252--262.
\newblock \href {https://doi.org/10.1016/j.commatsci.2018.01.014}
  {\path{doi:10.1016/j.commatsci.2018.01.014}}.

\bibitem{sun2019non}
Y.~Sun, J.~Luo, J.~Zhu, K.~Zhou, A non-isothermal phase field study of the
  shape memory effect and pseudoelasticity of polycrystalline shape memory
  alloys, Computational Materials Science 167 (2019) 65--76.
\newblock \href {https://doi.org/10.1016/j.commatsci.2019.05.036}
  {\path{doi:10.1016/j.commatsci.2019.05.036}}.

\bibitem{xu2020phase}
B.~Xu, G.~Kang, C.~Yu, Q.~Kan, {Phase field simulation on the grain size
  dependent super-elasticity and shape memory effect of nanocrystalline NiTi
  shape memory alloys}, International Journal of Engineering Science 156 (2020)
  103373.
\newblock \href {https://doi.org/10.1016/j.ijengsci.2020.103373}
  {\path{doi:10.1016/j.ijengsci.2020.103373}}.

\bibitem{xu2021phase}
B.~Xu, G.~Kang, {Phase field simulation on the super-elasticity, elastocaloric
  and shape memory effect of geometrically graded nano-polycrystalline NiTi
  shape memory alloys}, International Journal of Mechanical Sciences 201 (2021)
  106462.
\newblock \href {https://doi.org/10.1016/j.ijmecsci.2021.106462}
  {\path{doi:10.1016/j.ijmecsci.2021.106462}}.

\bibitem{wendler2017mesoscale}
F.~Wendler, H.~Ossmer, C.~Chluba, E.~Quandt, M.~Kohl, {Mesoscale simulation of
  elastocaloric cooling in SMA films}, Acta Materialia 136 (2017) 105--117.
\newblock \href {https://doi.org/10.1016/j.actamat.2017.06.044}
  {\path{doi:10.1016/j.actamat.2017.06.044}}.

\bibitem{hou2021effect}
X.~Hou, X.~Li, J.~Zhang, S.~P. Bag, H.~Li, J.~Wang, Effect of grain size on the
  electrocaloric properties of polycrystalline ferroelectrics, Physical Review
  Applied 15~(5) (2021) 054019.
\newblock \href {https://doi.org/10.1103/PhysRevApplied.15.054019}
  {\path{doi:10.1103/PhysRevApplied.15.054019}}.

\bibitem{moya2020caloric}
X.~Moya, N.~Mathur, Caloric materials for cooling and heating, Science
  370~(6518) (2020) 797--803.
\newblock \href {https://doi.org/10.1126/science.abb0973}
  {\path{doi:10.1126/science.abb0973}}.

\bibitem{yuan2019elastocaloric}
B.~Yuan, X.~Zhu, X.~Zhang, M.~Qian, Elastocaloric effect with small hysteresis
  in bamboo-grained {Cu--Al--Mn} microwires, Journal of Materials Science
  54~(13) (2019) 9613--9621.
\newblock \href {https://doi.org/10.1007/s10853-019-03592-8}
  {\path{doi:10.1007/s10853-019-03592-8}}.

\bibitem{chen2019giant}
H.~Chen, F.~Xiao, X.~Liang, Z.~Li, Z.~Li, X.~Jin, T.~Fukuda, Giant
  elastocaloric effect with wide temperature window in an al-doped
  nanocrystalline {Ti--Ni--Cu} shape memory alloy, Acta Materialia 177 (2019)
  169--177.
\newblock \href {https://doi.org/10.1016/j.actamat.2019.07.033}
  {\path{doi:10.1016/j.actamat.2019.07.033}}.

\bibitem{hou2018ultra}
H.~Hou, P.~Finkel, M.~Staruch, J.~Cui, I.~Takeuchi, Ultra-low-field
  magneto-elastocaloric cooling in a multiferroic composite device, Nature
  communications 9~(1) (2018) 1--8.
\newblock \href {https://doi.org/10.1038/s41467-018-06626-y}
  {\path{doi:10.1038/s41467-018-06626-y}}.

\bibitem{pataky2015elastocaloric}
G.~J. Pataky, E.~Ertekin, H.~Sehitoglu, Elastocaloric cooling potential of
  {NiTi}, {Ni$_2$FeGa}, and {CoNiAl}, Acta Materialia 96 (2015) 420--427.
\newblock \href {https://doi.org/10.1016/j.actamat.2015.06.011}
  {\path{doi:10.1016/j.actamat.2015.06.011}}.

\bibitem{qian2016elastocaloric}
S.~Qian, Y.~Geng, Y.~Wang, T.~E. Pillsbury, Y.~Hada, Y.~Yamaguchi, K.~Fujimoto,
  Y.~Hwang, R.~Radermacher, J.~Cui, et~al., Elastocaloric effect in {CuAlZn}
  and {CuAlMn} shape memory alloys under compression, Philosophical
  Transactions of the Royal Society A: Mathematical, Physical and Engineering
  Sciences 374~(2074) (2016) 20150309.
\newblock \href {https://doi.org/10.1098/rsta.2015.0309}
  {\path{doi:10.1098/rsta.2015.0309}}.

\bibitem{grunebohm2018origins}
A.~Gr{\"u}nebohm, Y.-B. Ma, M.~Marathe, B.-X. Xu, K.~Albe, C.~Kalcher, K.-C.
  Meyer, V.~V. Shvartsman, D.~C. Lupascu, C.~Ederer, Origins of the inverse
  electrocaloric effect, Energy Technology 6~(8) (2018) 1491--1511.
\newblock \href {https://doi.org/10.1002/ente.201800166}
  {\path{doi:10.1002/ente.201800166}}.

\bibitem{anderson1998diffuse}
D.~M. Anderson, G.~B. McFadden, A.~A. Wheeler, Diffuse-interface methods in
  fluid mechanics, Annual Review of Fluid Mechanics 30~(1) (1998) 139--165.
\newblock \href {https://doi.org/10.1146/annurev.fluid.30.1.139}
  {\path{doi:10.1146/annurev.fluid.30.1.139}}.

\bibitem{Khachaturyan1997Three}
A.~G. Khachaturyan, Y.~Wang, {Three-Dimensional Field Model and Computer
  Modeling of Martensitic Transformations}, Acta Materialia 45~(2) (1997)
  759--773.
\newblock \href {https://doi.org/10.1016/S1359-6454(96)00180-2}
  {\path{doi:10.1016/S1359-6454(96)00180-2}}.

\bibitem{artemev2001three}
A.~Artemev, Y.~M. Jin, A.~G. Khachaturyan, Three-dimensional phase field model
  of proper martensitic transformation, Acta Materialia 49~(7) (2001)
  1165--1177.
\newblock \href {https://doi.org/10.1016/S1359-6454(01)00021-0}
  {\path{doi:10.1016/S1359-6454(01)00021-0}}.

\bibitem{Kartha1995Disorder}
S.~Kartha, J.~A. Krumhansl, J.~P. Sethna, L.~K. Wickham, {Disorder-driven
  pretransitional tweed pattern in martensitic transformations}, Physical
  Review B 52~(2) (1995) 803--822.
\newblock \href {https://doi.org/10.1103/PhysRevB.52.803}
  {\path{doi:10.1103/PhysRevB.52.803}}.

\bibitem{yi2016real}
M.~Yi, B.-X. Xu, A real-space and constraint-free phase field model for the
  microstructure of ferromagnetic shape memory alloys, International Journal of
  Fracture 202~(2) (2016) 179--194.
\newblock \href {https://doi.org/10.1007/s10704-016-0152-4}
  {\path{doi:10.1007/s10704-016-0152-4}}.

\bibitem{gurtin1996generalized}
M.~E. Gurtin, {Generalized Ginzburg--Landau and Cahn--Hilliard equations based
  on a microforce balance}, Physica D: Nonlinear Phenomena 92~(3-4) (1996)
  178--192.
\newblock \href {https://doi.org/10.1016/0167-2789(95)00173-5}
  {\path{doi:10.1016/0167-2789(95)00173-5}}.

\bibitem{Landis2008A}
C.~M. Landis, {A continuum thermodynamics formulation for
  micro-magnetomechanics with applications to ferromagnetic shape memory
  alloys}, Journal of the Mechanics and Physics of Solids 56~(10) (2008)
  3059--3076.
\newblock \href {https://doi.org/10.1016/j.jmps.2008.05.004}
  {\path{doi:10.1016/j.jmps.2008.05.004}}.

\bibitem{nishiyama1958temperature}
Z.~Nishiyama, A.~Tsubaki, H.~Suzuki, Y.~Yamada, Temperature distribution during
  the martensite plate formation, Journal of the Physical Society of Japan
  13~(10) (1958) 1084--1090.
\newblock \href {https://doi.org/10.1143/jpsj.13.1084}
  {\path{doi:10.1143/jpsj.13.1084}}.

\bibitem{gurtin1966clausius}
M.~E. Gurtin, W.~O. Williams, On the clausius--duhem inequality, Zeitschrift
  f{\"u}r angewandte Mathematik und Physik ZAMP 17~(5) (1966) 626--633.
\newblock \href {https://doi.org/10.1007/BF01597243}
  {\path{doi:10.1007/BF01597243}}.

\bibitem{coleman1974thermodynamics}
B.~D. Coleman, W.~Noll, The thermodynamics of elastic materials with heat
  conduction and viscosity, in: The Foundations of Mechanics and
  Thermodynamics, Springer, 1974, pp. 145--156.

\bibitem{truesdell2004non}
C.~Truesdell, W.~Noll, The non-linear field theories of mechanics, in: The
  Non-linear Field Theories of Mechanics, Springer, 2004, pp. 1--579.
\newblock \href {https://doi.org/10.1007/978-3-662-10388-3_1}
  {\path{doi:10.1007/978-3-662-10388-3_1}}.

\bibitem{allen1979microscopic}
S.~M. Allen, J.~W. Cahn, A microscopic theory for antiphase boundary motion and
  its application to antiphase domain coarsening, Acta Metallurgica 27~(6)
  (1979) 1085--1095.
\newblock \href {https://doi.org/10.1016/0001-6160(79)90196-2}
  {\path{doi:10.1016/0001-6160(79)90196-2}}.

\bibitem{ohmer2022phase}
D.~Ohmer, M.~Yi, O.~Gutfleisch, B.-X. Xu, Phase-field modelling of paramagnetic
  austenite--ferromagnetic martensite transformation coupled with mechanics and
  micromagnetics, International Journal of Solids and Structures 238 (2022)
  111365.
\newblock \href {https://doi.org/10.1016/j.ijsolstr.2021.111365}
  {\path{doi:10.1016/j.ijsolstr.2021.111365}}.

\bibitem{Zhang2005Phase}
J.~X. Zhang, L.~Q. Chen, {Phase-field model for ferromagnetic shape-memory
  alloys}, Philosophical Magazine Letters 85~(10) (2005) 533--541.
\newblock \href {https://doi.org/10.1080/09500830500385527}
  {\path{doi:10.1080/09500830500385527}}.

\bibitem{daly2007stress}
S.~Daly, G.~Ravichandran, K.~Bhattacharya, {Stress-induced martensitic phase
  transformation in thin sheets of Nitinol}, Acta Materialia 55~(10) (2007)
  3593--3600.
\newblock \href {https://doi.org/10.1016/j.actamat.2007.02.011}
  {\path{doi:10.1016/j.actamat.2007.02.011}}.

\bibitem{shimizu1982crystallographic}
K.~Shimizu, Y.~Okumura, H.~Kubo, Crystallographic and morphological studies on
  the fcc to fct transformation in {Mn--Cu} alloys, Transactions of the Japan
  Institute of Metals 23~(2) (1982) 53--59.
\newblock \href {https://doi.org/10.2320/matertrans1960.23.53}
  {\path{doi:10.2320/matertrans1960.23.53}}.

\bibitem{manosa2017materials}
L.~Ma{\~n}osa, A.~Planes, Materials with giant mechanocaloric effects: cooling
  by strength, Advanced Materials 29~(11) (2017) 1603607.
\newblock \href {https://doi.org/10.1002/adma.201603607}
  {\path{doi:10.1002/adma.201603607}}.

\bibitem{yeddu2012three-dimensional}
H.~K. Yeddu, A.~Malik, J.~Agren, G.~Amberg, A.~Borgenstam, Three-dimensional
  phase-field modeling of martensitic microstructure evolution in steels, Acta
  Materialia 60~(4) (2012) 1538--1547.
\newblock \href {https://doi.org/10.1016/j.actamat.2011.11.039}
  {\path{doi:10.1016/j.actamat.2011.11.039}}.

\bibitem{de1993crossover}
C.~S. De~Melo, M.~Randeria, J.~R. Engelbrecht, {Crossover from BCS to Bose
  superconductivity: Transition temperature and time-dependent Ginzburg-Landau
  theory}, Physical Review Letters 71~(19) (1993) 3202.
\newblock \href {https://doi.org/10.1103/physrevlett.71.3202}
  {\path{doi:10.1103/physrevlett.71.3202}}.

\bibitem{man2010microstructural}
J.~Man, J.~Zhang, Y.~Rong, {Microstructural evolution of Mn-rich
  antiferromagnetic Mn--Cu alloy under temperature field}, Applied Physics
  Letters 96~(13) (2010) 131904.
\newblock \href {https://doi.org/10.1063/1.3378810}
  {\path{doi:10.1063/1.3378810}}.

\bibitem{Dhote2012Dynamic}
R.~P. Dhote, R.~V. Melnik, J.~Zu, {Dynamic thermo-mechanical coupling and size
  effects in finite shape memory alloy nanostructures}, Computational Materials
  Science 63 (2012) 105--117.
\newblock \href {https://doi.org/10.1016/j.commatsci.2012.05.060}
  {\path{doi:10.1016/j.commatsci.2012.05.060}}.

\bibitem{jacobs2003simulations}
A.~Jacobs, S.~Curnoe, R.~Desai, Simulations of cubic-tetragonal ferroelastics,
  Physical Review B 68~(22) (2003) 224104.

\bibitem{yi2014constraint}
M.~Yi, B.-X. Xu, A constraint-free phase field model for ferromagnetic domain
  evolution, Proceedings of the Royal Society A: Mathematical, Physical and
  Engineering Sciences 470~(2171) (2014) 20140517.
\newblock \href {https://doi.org/10.1098/rspa.2014.0517}
  {\path{doi:10.1098/rspa.2014.0517}}.

\bibitem{tonks2012object}
M.~R. Tonks, D.~Gaston, P.~C. Millett, D.~Andrs, P.~Talbot, An object-oriented
  finite element framework for multiphysics phase field simulations,
  Computational Materials Science 51~(1) (2012) 20--29.
\newblock \href {https://doi.org/10.1016/j.commatsci.2011.07.028}
  {\path{doi:10.1016/j.commatsci.2011.07.028}}.

\bibitem{Artemev2000Phase}
A.~Artemev, A.~G. Khachaturyan, {Phase field model and computer simulation of
  martensitic transformation under applied stresses}, Materials Science Forum
  327 (2000) 347--350.

\bibitem{alvarez2017conventional}
P.~{\'A}lvarez-Alonso, C.~O. Aguilar-Ortiz, E.~Villa, A.~Nespoli,
  H.~Flores-Z{\'u}{\~n}iga, V.~Chernenko, Conventional and inverse
  elastocaloric effect in {Ni-Fe-Ga} and {Ni-Mn-Sn} ribbons, Scripta Materialia
  128 (2017) 36--40.
\newblock \href {https://doi.org/10.1016/j.scriptamat.2016.09.033}
  {\path{doi:10.1016/j.scriptamat.2016.09.033}}.

\bibitem{Xiao2021}
F.~Xiao, Z.~Li, H.~Chen, X.~Jin, A.~Planes, T.~Fukuda, {Origin of the inverse
  elastocaloric effect in a Ni-rich Ti-Ni shape memory alloy induced by
  oriented nanoprecipitates}, Physical Review Materials 5~(5) (2021) 1--10.
\newblock \href {https://doi.org/10.1103/PhysRevMaterials.5.053603}
  {\path{doi:10.1103/PhysRevMaterials.5.053603}}.

\bibitem{chen2021improved}
J.~Chen, L.~Xing, G.~Fang, L.~Lei, W.~Liu, {Improved elastocaloric cooling
  performance in gradient-structured NiTi alloy processed by localized laser
  surface annealing}, Acta Materialia 208 (2021) 116741.
\newblock \href {https://doi.org/10.1016/j.actamat.2021.116741}
  {\path{doi:10.1016/j.actamat.2021.116741}}.

\bibitem{Cheng2020}
X.~Cheng, Y.~Li, D.~Zhu, M.~Li, M.~Feng, {Effects of uniaxial compressive
  stress on the electrocaloric effect of ferroelectric ceramics}, Journal of
  Materials Science 55~(21) (2020) 8802--8813.
\newblock \href {https://doi.org/10.1007/s10853-020-04640-4}
  {\path{doi:10.1007/s10853-020-04640-4}}.

\bibitem{xu2016giant}
S.~Xu, H.-Y. Huang, J.~Xie, S.~Takekawa, X.~Xu, T.~Omori, R.~Kainuma, Giant
  elastocaloric effect covering wide temperature range in columnar-grained
  {Cu$_{71. 5}$Al$_{17. 5}$Mn$_{11}$} shape memory alloy, APL Materials 4~(10)
  (2016) 106106.
\newblock \href {https://doi.org/10.1063/1.4964621}
  {\path{doi:10.1063/1.4964621}}.

\end{thebibliography}

\appendix

\section{Rotation matrix}\label{A1}

\begin{equation}
\underline{\mathbf{K}}_\sigma=
\left(             
  \begin{array}{cccccc}   
    Q_{11}^2 & Q_{12}^2 & Q_{13}^2 & 2Q_{11}Q_{12} & 2Q_{12}Q_{13}& 2Q_{11}Q_{13}\\  
    Q_{21}^2 & Q_{22}^2 & Q_{23}^2 & 2Q_{21}Q_{22} & 2Q_{22}Q_{23}& 2Q_{21}Q_{23}\\  
    Q_{31}^2 & Q_{32}^2 & Q_{33}^2 & 2Q_{31}Q_{32} & 2Q_{33}Q_{32}& 2Q_{31}Q_{33}\\
    Q_{11}Q_{21} & Q_{12}Q_{22} & Q_{13}Q_{23} & Q_{11}Q_{22}+Q_{12}Q_{21} & Q_{12}Q_{23}+Q_{13}Q_{22}& Q_{11}Q_{23}+Q_{13}Q_{21}\\
    Q_{21}Q_{31} & Q_{22}Q_{32} & Q_{23}Q_{33} & Q_{21}Q_{32}+Q_{22}Q_{31} & Q_{22}Q_{33}+Q_{23}Q_{32}& Q_{21}Q_{33}+Q_{23}Q_{31}\\
    Q_{31}Q_{11} & Q_{12}Q_{32} & Q_{13}Q_{33} & Q_{31}Q_{12}+Q_{32}Q_{11} & Q_{32}Q_{13}+Q_{33}Q_{12}& Q_{31}Q_{13}+Q_{11}Q_{33}\\
  \end{array}
\right)
\end{equation}

\end{document}